\documentclass[times]{asjcauth}
\usepackage{url}
\usepackage{microtype}
\newlength\figwidth
\usepackage{amsfonts}
\usepackage{mathtools}
\usepackage{mathrsfs}
\usepackage{xspace}
\usepackage{amsmath}
\usepackage{amssymb}
\usepackage[compress ]{cite}
\usepackage{float}
\usepackage{dblfloatfix}
\newcommand{\hinf}{\ensuremath{\mathscr{H}_\infty}\xspace}
\newcommand{\bi}[1]{\boldsymbol{\mathit{#1}}}
\DeclareMathAlphabet{\mathcalligra}{T1}{calligra}{m}{n}
\newtheorem{lemma}{Lemma}
\newtheorem{theorem}{Theorem}
\theoremstyle{definition}
\newtheorem{definition}{Definition}
\newtheorem{remark}{Remark}
\newtheorem{assumption}{Assumption}
\newenvironment{proof}{\paragraph{Proof:}}{\hfill$\square$}

\begin{document}
\title{ROBUST ADAPTIVE SLIDING MODE CONTROL OF MARKOVIAN JUMP SYSTEMS WITH UNCERTAIN MODE-DEPENDENT TIME-VARYING DELAYS AND PARTLY UNKNOWN TRANSITION PROBABILITIES}
\runninghead{N.~Zohrabi et al.: Sliding Control of Mode-dep. Delayed Unknown MJS}
\author{Nasibeh Zohrabi,  Hasan Zakeri, Amir Hossein Abolmasoumi, Hamid Reza Momeni} 
\address{N. Zohrabi is with Department of Electrical and Computer Engineering, Mississippi State University, US.\\\url{nz75@msstate.edu}\\
H. Zakeri is with Department of Electrical Engineering, University of Notre Dame, US. \\\url{hzakeri@nd.edu}\\
A. H. Abolmasoumi is with Department of Electrical Engineering, Arak University, Arak, Iran. \\\url{a-abolmasoumi@araku.ac.ir}\\
H. R. Momeni is with Department of Electrical Engineering, Tarbiat Modares University, Tehran, Iran.\\ \url{momeni_h@modares.ac.ir}
}

\begin{abstract}
This paper deals with the problems of stochastic stability and sliding mode control for a class of continuous-time Markovian jump systems with mode-dependent time-varying delays and partly unknown transition probabilities. The design method is general enough to cover a wide spectrum of systems from those with completely known transition probability rates to those with completely unknown transition probability rates. Based on some mode-dependent Lyapunov-Krasovski functionals and making use of the free-connection weighting matrices, new delay-dependent conditions guaranteeing the existence of linear switching surfaces and the stochastic stability of sliding mode dynamics are derived in terms of linear matrix inequalities (LMIs). Then, a sliding mode controller is designed such that the resulted closed-loop system's trajectories converge to predefined sliding surfaces in a finite time and remain there for all subsequent times. This paper also proposes an adaptive sliding mode controller design method which applies to cases in which mode-dependent time-varying delays are unknown. All the conditions obtained in this paper are in terms of LMI feasibility problems. Numerical examples are given to illustrate the effectiveness of the proposed methods.
\end{abstract}

\maketitle
\thispagestyle{empty}
\makeatletter
\def\@oddfoot{}
\def\@evenfoot{}
\let\@evenhead\@oddhead
\makeatother

\section{Introduction}\label{sec:intro}
Markovian jump systems (MJSs), first introduced in~\cite{bib33}, are a class of stochastic hybrid systems described by a set of classical differential equations along with a finite state Markov process representing the discrete state or jump. Transition probability rates are statistical values determining the behavior of system's jumps. The complete knowledge of the transition probabilities simplifies the analysis and control of the MJSs to a large degree. Due to vast applications in various real world problems, including those in the networked control systems, aerospace systems, and manufacturing systems (see~\cite{bib2,bib3,bib4,bib5, jadid1,Z3, Z4}, etc.), in the past few decades many have devoted their research to the study of Markovian jump systems and several results have been achieved.  For example,  see~\cite{bib7, bib100,bib101,bib11} and references therein. However, as a drawback, these results  suffer from the assumption of fully-known transition probability rates.

{Despite this common assumption,} in most cases, all or parts of the elements in the transition probabilities matrix are not known a priori. The likelihood of a complete measurement regarding transition probabilities in practical cases is quite controversial, and it can also simultaneously be costly or time-consuming. Therefore, rather than gauging or estimating all the elements of transition probabilities matrix, it is a better choice to study more general MJSs with partly unknown transition probabilities. Recently, several interesting results on stability, stabilisation and filtering problems for MJSs with partly unknown transition probabilities have been addressed. For example, we refer readers to~\cite{jadid2, bib13,bib103,bib19,bib21}.

Meanwhile, time delays occur frequently in many practical control systems such as biological systems, heating systems and networked control systems. Particularly, time delays are well known as a source of instability and poor performance of a control system~\cite{bib49}. Accordingly, many results related to stability, stabilisation, and filtering of time delay Markovian jump systems have been obtained. See ~\cite{bib49,bib50,bib51,bib52,newbib1} and references therein for example. In terms of their stability conditions, these results are mainly classified into two categories: delay-dependent and delay-independent conditions. Applying the information regarding the size of delays, the delay-dependent criteria are considered to be less-conservative than the delay-independent ones, especially when the size of the delay is small. Recently, Markovian jump systems with mode-dependent time delays where the time delays depend on the system modes have been studied, and many topics such as stability, stabilisation and control of such systems have been investigated~\cite{bib56,bib55,bib102, jadid3}. In this paper, the mode-dependent time-varying delayed Markovian jump system is considered, and new delay-dependent conditions are obtained in terms of less-conservative LMIs.
 
On the other hand, sliding mode control (SMC) is one of the most important robust control methods for uncertain or nonlinear systems. The main concept of SMC design is to utilize a discontinuous control law to drive the state trajectories of the closed-loop system to the  predesigned sliding surface in a finite time and to maintain there for all subsequent times. The sliding surface is designed in advance with desired properties such as stability, regulation, disturbance rejection capability, tracking, etc. {During the last decade, sliding mode control for MJSs has attracted a considerable interest.  See~\cite{bib23,bib24, jadid4, bib25,bib26,Z1, Z2}  for example.} However, in most of the mentioned works, it is assumed that all elements of transition probability rate matrix are known and accessible. This assumption drastically eases up the design process but at the same time sets limits on the generality of results especially in practical applications. 

Authors in~\cite{IFAC} investigate the problem of sliding mode control for Markovian jump systems with partly unknown transition probabilities, however they do not address the delay problem. Due to the significant effects of delays on the system's performance, it is essential to consider potential delays in the study of control and stochastic stability. Thus, bringing the mode-dependent time-varying delay in the problem of MJS sliding mode design with partly known probability rates could be considered as one of the main contributions of this paper.   As a result of taking the delay into account, the obtained stochastic stability conditions using the LMI framework would become  much more complex  than those presented in the earlier works in this field, such as in~\cite{IFAC}.  

{In most practical situations, the time delay functions are not exactly known though, in some cases, their bounds are available. Therefore, the desirable delay states cannot be employed in the SMC law in these cases. However, in some results in the literature such as~~\cite{bib555,bib666}, not only are the bounds of the time delays assumed to be known, but the time delay functions used in control law are also supposed to be known precisely. This is definitely a very restrictive condition. To overcome this problem, here we present a new adaptive sliding mode controller for MJSs with unknown mode-dependent time-varying delays.}

Motivated by the above discussion, this paper considers the SMC design for delayed Markovian jump systems with partly unknown transition rates.  The stochastic stability of sliding mode dynamics is assured based on a new stochastic Lyapunov-Krasovskii functional combining with Jensen's inequality and usage of free-conn\-ec\-tion weighting matrices. The Lyapunov functional includes an upper bound, a lower bound, and a derivative bound of the mode-dependent time-varying delay, so less-conservative delay-dependent conditions are obtained in terms of LMIs, guaranteeing the existence of the desired linear sliding surface and the stochastic stability of sliding mode dynamics. Afterward, with the assistance of a mode-dependent Lyapunov function and free weighting connection matrices, we design a sliding mode controller to ensure the reachability of closed-loop's trajectories to the desired switching surface in a finite time. Finally, we propose a novel adaptive SMC law design method to handle unknown mode-dependent time-varying delays in the system under consideration. This is another major contribution of this paper which is presented in Theorem~\ref{thrm3}. 

The remainder of this paper is organized as follows: Section~\ref{sec:problem} gives the problem statement and preliminary information. Section~\ref{sec:main} first considers the problems of stochastic stability of sliding mode dynamics and the design procedure of a desired SMC law to ensure the stochastic stability of closed-loop system. {Afterward, it generalizes the results by proposing an adaptive SMC law.} Numerical examples and the conclusion are given in sections~\ref{sec:numerical} and~\ref{sec:conc}, respectively. 
\section{Problem statement and preliminaries}\label{sec:problem}
Consider the following stochastic continuous-time Markovian jump system with mode-dependent time-varying delays defined in the probability space \((\Omega,\mathcal{F},\mathbb{P})\):	
\begin{equation}\label{eq:sys1}
\begin{aligned}
\bi{\dot {x}}(t)=\ &\bi A(r_t)\bi x(t)+\bi{A_d}(r_t)\bi x(t-\tau_{r_t}(t))\\
&+\bi B(r_t)[\bi u(t)+\bi F(r_t)\bi w(t)]
\end{aligned}
\end{equation}
where \(\bi x(t)\in{\mathbb R^n}\) is the state vector, \(\bi u(t)\in{\mathbb R^m}\) is controller input, \(\bi w(t)\in{\mathbb R^l}\)  is the disturbance. \(\{r_t,t>0\}\)   is the continuous-time Markov process which takes value in a finite state space \(\ell=\{1,2,...,N\}\) with generator \(\lambda_{ij}\),
\begin{equation}\label{generator}
\textup{Pr}\left(r_{t+h}=j\mid r_t=i\right)=
	\begin{cases}
		\lambda_{ij}h+o(h),&\text{ if }j\neq i\\
		1+\lambda_{ii}h+o(h),&\text{ if }j=i
	\end{cases}
\end{equation}
where \(\lambda_{ij}\geq0\  (i,j\in{\ell},j\neq{i})\)  represents the transition rate from mode \(i\)   at time \(t\)  to mode \(j\)  at time \(t+h\) with \(\lambda_{ii}=-\sum_{j=1,j\neq{i}}^{N}\lambda_{ij}\) for each \(i\in{\ell},\)   and \(h>0,\lim_{h\to0}{(o(h)/h)=0}\). Besides, the Markov process transition probability rate matrix \(\bi{\Lambda}\)  is defined by
\begin{equation*}
\bi \Lambda=\left[\begin{array}{llcl}
\lambda_{11}&\lambda_{12}&{\dots}&\lambda_{1N}\\
\lambda_{21}&\lambda_{22}&{\dots}&\lambda_{2N}\\
\ \vdots&\ \vdots&\ddots&\ \vdots\\
\lambda_{N1}&\lambda_{N2}&{\dots}&\lambda_{NN}
\end{array}\right]
\end{equation*}
For convenience, for each possible value  \(r_t=i,\ i\in{\ell}\), we define \(\bi A(r_t)=\bi{A_i},\ \bi{A_d}(r_t)=\bi{A_{di}},\ \bi B(r_t)=\bi{B_i}\)  and \(\bi F(r_t)=\bi{F_i}\). Then, system~\eqref{eq:sys1} can be described by
\begin{multline}\label{eq:systemxdot}
\bi{\dot {x}}(t)=\bi{A_i}\bi x(t)\\+\bi{A_{di}}\bi x(t-\tau_i(t))+\bi{B_i}[\bi u(t)+\bi{F_i}\bi w(t)]
\end{multline}
where \(\bi {A_i},\ \bi{A_{di}},\ \bi {B_i}\) and  \(\bi {F_i}\)  are known constant matrices of appropriate dimensions. It is assumed that
\begin{equation}
{\Vert \bi{F_i}\bi w(t)\Vert}\leq f_i,\quad\quad i\in{\ell}
\end{equation}
with \(f_i>0\). Besides, \(\tau_i(t)\) denotes mode-dependent time-varying delay {(whether known or unknown)}, satisfying the following conditions:
\begin{equation}\label{delay}
0\leq h_1\leq h_{1i}\leq \tau_i(t)\leq h_{2i}\leq h_2,\qquad \dot\tau_i(t)\leq\mu_i
\end{equation}
where \(h_1=\min_{\ i\in\ell} h_{1i}\) and \(h_2=\max_{\ i\in\ell} h_{2i}\). In this paper, the transition probability rates are considered to be partly unknown, i.e., some elements in matrix \(\bi\Lambda\) are unknown (They can be fully known or fully unknown as well). For distinctive notation, we define \(\ell=\ell_\mathcal K^i\cup\ell_{u\mathcal K}^i\) by:
\begin{equation}\label{lambda}
	\begin{aligned}
		\ell_\mathcal K^i&\triangleq\{j:\lambda_{ij}\text{ is known}\}\\
		\ell_{u\mathcal K}^i&\triangleq\{j:\lambda_{ij}\text{ is unknown}\}\\
	\end{aligned}
\end{equation}
and if \(\ell_\mathcal K^i\neq\emptyset\), it is also described as
\begin{equation}\label{known tr}
\ell_\mathcal K^i=\left(\kappa_1^i,\dots,\kappa_q^i\right),\quad 1\leq q\leq N
\end{equation}
where \(\kappa_q^i\in{\mathbb{N}^+}\)  stands for the \(q\)\/\textbf{th} known element with index  \(\kappa_q^i\)   in the \(i\mathbf{th}\) row of matrix \(\bi\Lambda\).  Taking these definitions into account, we study a more general class of Markovian jump systems.

Before proceeding further, we will introduce the following definition and some lemmas which are indispensable in deriving the proposed stability criterion.
\begin{definition}
\cite{bib7} The Markovian jump system  \(\bi{\dot x}(t)=\bi A(r_t)\bi x(t)\) is said to be stochastically stable~(SS) if there exists a finite positive constant \(T(\bi{x_0},r_0)\)  such that the following holds for any initial condition  \((\bi{x_0},r_0)\):
\[
\mathsf{E}\left\{\int_0^\infty\Vert \bi x(t)\Vert^2dt\mid \bi{x_0},r_0\right\}<T(\bi{x_0},r_0)
\]
\end{definition}

\begin{lemma}\label{lem5}
\cite{bib34}
Let \(\bi A,\bi D\) and \(\bi F\) be real matrices of appropriate dimensions with \(\bi F\) satisfying \(\bi F^T\bi F<\bi I\). Then for any scalar \(\epsilon>0\) and vectors \(\bi x,\bi y\in \mathbb R^n\), the following statement holds:
\begin{equation}
2\bi x^T\bi A\bi F\bi D\bi y\leq \epsilon^{-1}\bi x^T\bi A\bi A^T\bi x+\epsilon \bi y^T\bi D^T\bi D\bi y.
\end{equation}
\end{lemma} 

\begin{lemma}\label{lem12}
\cite{bib37}
Suppose \(\gamma_1\leq \gamma(t)\leq \gamma_2\), where \(\gamma(.): \mathbb R_+\) (or\(~\mathbb Z_+) \rightarrow \mathbb R_+\) (or \(\mathbb Z_+)\). Then, for any constant matrices \(\bi{\Xi_1},\bi{\Xi_2}\) and \(\bi\Xi\) with proper dimensions, the following matrix inequality 
\begin{equation}
\bi\Xi+\left(\gamma(t)-\gamma_1\right)\bi{\Xi_1}+\left(\gamma_2-\gamma(t)\right)\bi{\Xi_2}<0
\end{equation}
holds, if and only if
\begin{equation}
\begin{aligned}
&\bi\Xi+\left(\gamma_2-\gamma_1\right)\bi{\Xi_1}<0,\\
&\bi\Xi+\left(\gamma_2-\gamma_1\right)\bi{\Xi_2}<0.
\end{aligned}
\end{equation}
\end{lemma}

\section{Main results}\label{sec:main}
In order to obtain a regular form of system~\eqref{eq:systemxdot}, we first choose a nonsingular matrix \(\bi{T_i}\) such that following equality holds~\cite{bib23}:
\begin{align*}
\bi{T_i}\bi{B_i}&=\left[\begin{array}{l}
\bi 0_{(n-m)\times m}\\
\bi{B_{2i}}
\end{array}\right]
\intertext{in which \(\bi{B_{2i}}\in{\mathbb R^{m\times m}}\)  is nonsingular. For convenience, we partition \(\bi{T_i}\)  as follows:}
\bi{T_i}&=\left[\begin{array}{l}
\bi{U_{2i}}^T\\
\bi{U_{1i}}^T
\end{array}\right]
\end{align*}
where \(\bi{U_{1i}}\in{\mathbb R^{n\times m}}\)  and \(\bi{U_{2i}}\in{\mathbb R^{n\times(n-m)}}\)  are two sub-blocks of a unitary matrix resulting from the singular value decomposition of \(\bi{B_i}\) , that is:
\begin{equation*}
\bi{B_i}=\left[\begin{array}{ll}
\bi{U_{1i}}&\bi{U_{2i}}
\end{array}\right]
\left[\begin{array}{l}
\bi{\Sigma_i}\\
\bi 0_{(n-m)\times m}
\end{array}\right]
\bi{J_i}^T
\end{equation*}
where \(\bi{\Sigma_i}\in{\mathbb R^{m\times m}}\)  is a diagonal positive-definite matrix and \(\bi{J_i}\in{\mathbb R^{m\times m}}\)  is a unitary matrix. Then the state transformation \(\bi z(t)=\bi{T_i}\bi x(t)\) is applied to system~\eqref{eq:systemxdot} to derive the following regular form:
\begin{equation}\label{regular}
\begin{aligned}
\bi{\dot z}(t)= \ &\bi{\bar A_i}\bi z(t)+\bi{\bar A_{di}}\bi z(t-\tau_i(t))\\
&+ \left[\begin{array}{l}
\bi 0_{(n-m)\times m}\\
\bi{B_{2i}}
\end{array}\right][\bi u(t)+\bi{F_i}\bi w(t)]
\end{aligned}
\end{equation}
in which, \(\bi{\bar A_i}=\bi{T_i}\bi{A_i}\bi{T_i}^{-1}\) and \(\bi{\bar A_{di}}=\bi{T_i}\bi{A_{di}}\bi{T_i}^{-1}\). system~\eqref{regular} can be written as follows:
\begin{equation}
\begin{aligned}
\bi{\dot z_1}(t)=\ &\bi{\bar A_{11i}}\bi{z_1}(t)+\bi{\bar A_{12i}}\bi{z_2}(t)+\bi{\bar A_{d11i}}\bi{z_1}(t-\tau_i(t))\\
&+\bi{\bar A_{d12i}}\bi{z_2}(t-\tau_i(t))\label{systemz1}
\end{aligned}
\end{equation}
\begin{equation}
\begin{aligned}
\bi{\dot z_2}(t)=\ &\bi{\bar A_{21i}}\bi{z_1}(t)+\bi{\bar A_{22i}}\bi{z_2}(t)+\bi{\bar A_{d21i}}\bi{z_1}(t-\tau_i(t))\\
&+\bi{\bar  A_{d22i}}\bi{z_2}(t-\tau_i(t))+\bi{B_{2i}}[\bi u(t)+\bi{F_i}\bi w(t)]\label{systemz2}
\end{aligned}
\end{equation}
where \(\bi{z_1}(t)\in{\mathbb R^{n-m}},\bi{z_2}(t)\in{\mathbb R^{m}}\)  and other parameters are obtained as follows:
\begin{align*}
\bi{\bar A_{11i}}&=\bi{U_{2i}}^T\bi{A_i}\bi{U_{2i}},\ \bi{\bar A_{12i}}=\bi{U_{2i}}^T\bi{A_i}\bi{U_{1i}},\\ 
\bi{\bar A_{21i}}&=\bi{U_{1i}}^T\bi{A_i}\bi{U_{2i}},\ \bi{\bar A_{22i}}=\bi{U_{1i}}^T\bi{A_i}\bi{U_{1i}},\\
\bi{\bar A_{d11i}}&=\bi{U_{2i}}^T\bi{A_{di}}\bi{U_{2i}},\ \bi{\bar A_{d12i}}=\bi{U_{2i}}^T\bi{A_{di}}\bi{U_{1i}},\\
\bi{\bar A_{d21i}}&=\bi{U_{1i}}^T\bi{A_{di}}\bi{U_{2i}},\ \bi{\bar A_{d22i}}=\bi{U_{1i}}^T\bi{A_{di}}\bi{U_{1i}},\\
\bi{B_{2i}}&=\bi{\Sigma_i}\bi{J_i}^T.
\end{align*}
Based on sliding mode control theory~\cite{bib28,bib29}, it is known that~\eqref{systemz1} denotes the sliding mode dynamics. Therefore, we design the following linear sliding surface:
\begin{equation}\label{sliding surface}
\bi s(t)=\left[\begin{array}{ll}
\bi{C_{1i}}&\bi{C_{2i}}
\end{array}\right]\bi z(t)
\end{equation}
where \(\bi{C_{2i}}\)  is invertible for each \(i\in\ell\). By defining \(\bi{C_i}=\bi{C_{2i}}^{-1}\bi{C_{1i}}\) and 
substituting  \(\bi{z_2}(t)=-\bi{C_i}\bi{z_1}(t)\) and \(\bi{z_2}(t-\tau_i(t))=-\bi{C_i}\bi{z_1}(t-\tau_i(t))\) to sliding dynamics~\eqref{systemz1}, we have
\begin{equation}\label{sl}
\bi{\dot z_1}(t)=\bi{\tilde A_i}\bi{z_1}(t)+\bi{\tilde A_{di}}\bi{z_1}(t-\tau_i(t)),
\end{equation}
\begin{equation*}
\bi{\tilde A_i}=\bi{\bar A_{11i}}-\bi{\bar A_{12i}}\bi{C_i},\quad \bi{\tilde A_{di}}=\bi{\bar A_{d11i}}-\bi{\bar A_{d12i}}\bi{C_i}.
\end{equation*}
By means of sliding mode control theory, when the state trajectories of the closed-loop system drive onto the sliding surface and maintain there for all subsequent times, we have \(\bi{s}(t)=0\)  and \(\bi{\dot s}(t)=0\). Now, we are in the position to present main results of this paper. In the following, in Theorem~\ref{thrm1}, we design linear sliding surface parameter \(\bi{C_i}\)  for the stochastic stability of sliding mode dynamics~\eqref{sl}. Then, in Theorem~\ref{thrm2}, we construct a desired SMC law \(\bi u(t)\) which ensures that state trajectories of the closed-loop system enter the predefined sliding surface in finite time.
\begin{theorem}\label{thrm1}
The sliding mode dynamics~\eqref{sl} with mode-dependent time-varying delays \(\tau_i(t)\) and partly unknown transition probabilities~\eqref{lambda}, is stochastically stable if there exist matrices \(\bi{X_i} > 0,\bi{\hat Q_{1i}}>0,\bi{\hat Q_{2i}}>0,\bi{\hat Q_{3i}}>0,\bi{\hat Q_{1}}>0,\bi{\hat Q_{2}}>0,\bi{\hat Q_{3}}>0,\bi{\hat R_{1}}>0,\bi{\hat R_{2}}>0,\bi{V_i} =\bi{V_i}^T,\bi{\hat W_{ri}}=\bi{\hat W}^T_{\bi{ri}}\) with \(r=1,2,3\), \(\bi{\hat M_i},\ \bi{\hat N_i},\ \bi{\hat S_i}\), and \(\bi{Y_i}\)  {such that the sets of LMIs~\eqref{LMI1}-\eqref{LMI13} hold for each \(i\in\ell\).}
\begin{figure*}[h]
\begin{equation}\label{LMI1}
\bi{\hat\Delta_{1i}}=\begin{bmatrix}
\bi{\hat\theta_{bi}}+\bi{\hat\phi_i}+\bi{\hat\phi_i}^T&h_2\bi{\hat A_{im}}^T&h_{21}\bi{\hat A_{im}}^T&h_1\bi{\hat M_i}&h_{21}\bi{\hat N_i}&\bi{\Gamma_i}(\bi{X_i})\\
*&-h_2\bi{\hat R_1}&\bi 0&\bi 0&\bi 0&\bi 0\\
*&\bi 0&-h_{21}\bi{\hat R_2}&\bi 0&\bi 0&\bi 0\\
*&*&*&\mathllap{h_1}(\bi{\hat R_1}-2\bi{X_i})&\bi 0&\bi 0\\
*&*&*&\bi 0&\mathllap{{h_{21}}}\left(\bi{\hat R_1}+\bi{\hat R_2}-4\bi{X_i}\right)&\bi 0\\
*&*&*&*&*&-\bi{\Xi_i}(\bi{X_i})
\end{bmatrix}<0
\end{equation}
\begin{equation}\label{LMI2}
\bi{\hat\Delta_{2i}}=\begin{bmatrix}
\bi{\hat\theta_{bi}}+\bi{\hat\phi_i}+\bi{\hat\phi_i}^T&h_2\bi{\hat A_{im}}^T&h_{21}\bi{\hat A_{im}}^T&h_2\bi{\hat M_i}&h_{21}\bi{\hat S_i}&\bi{\Gamma_i}(\bi{X_i})\\
*&-h_2\bi{\hat R_1}&\bi 0&\bi 0&\bi 0&\bi 0\\
*&\bi 0&-h_{21}\bi{\hat R_2}&\bi 0&\bi 0&\bi 0\\
*&*&*&h_2(\bi{\hat R_1}-2\bi{X_i})&\bi 0&\bi 0\\
*&*&*&\bi 0&\mathllap{h_{21}}\left(\bi{\hat R_2}-2\bi{X_i}\right)&\bi 0\\
*&*&*&*&*&-\bi{\Xi_i}(\bi{X_i})
\end{bmatrix}<0
\end{equation}
\centering{\rule{0.8\linewidth}{0.4pt}}
\end{figure*}

\begin{equation}\label{LMI3}
\begin{bmatrix}
-\bi{V_i}&\bi{X_i}&\\
\bi{X_i}&-\bi{X_j}
\end{bmatrix}\leq0,\quad i\in{\ell_\mathcal K^i},\quad j\in{\ell_{u\mathcal K}^i}
\end{equation}

\begin{equation}\label{LMI4}
\bi{X_j}-\bi{V_i}\geq0,\quad i\in{\ell_{u\mathcal K}^i},\quad j=i
\end{equation}

\begin{equation}\label{LMI5}
\begin{bmatrix}
\lambda_{ij}\bi{\hat {Q_{1j}}}-2\bi{X_j}+\alpha\bi{\hat Q_1}&\lambda_{ij}\bi{X_j}\\
\lambda_{ij}\bi{X_j}&\lambda_{ij}\bi{\hat W_{1i}}
\end{bmatrix}<0, j\in{\ell_\mathcal K^i},\ j\neq i
\end{equation}

\begin{equation}\label{LMI6}
\bi{\hat Q_{1j}}-2\bi{X_j}+\bi{\hat W_{1i}}\leq0,\quad j\in{\ell_{u\mathcal K}^i},\quad j\neq i
\end{equation}

\begin{equation}\label{LMI7}
\bi{\hat Q_{1i}}-2\bi{X_i}+\bi{\hat W_{1i}}\geq0,\quad j=i
\end{equation}
\begin{equation}\label{LMI8}
\begin{bmatrix}
\lambda_{ij}\bi{\hat Q_{2j}}-2\bi{X_j}+\alpha\bi{\hat Q_2}&\lambda_{ij}\bi{X_j}\\
\lambda_{ij}\bi{X_j}&\lambda_{ij}\bi{\hat W_{2i}}
\end{bmatrix}<0, j\in{\ell_\mathcal K^i},\ j\neq i
\end{equation}

\begin{equation}\label{LMI9}
\bi{\hat Q_{2j}}-2\bi{X_j}+\bi{\hat W_{2i}}\leq0,\quad j\in{\ell_{u\mathcal K}^i},\quad j\neq i
\end{equation}

\begin{equation}\label{LMI10}
\bi{\hat Q_{2i}}-2\bi{X_i}+\bi{\hat W_{2i}}\geq0,\quad j=i
\end{equation}
\begin{equation}\label{LMI11}
\begin{bmatrix}
\lambda_{ij}\bi{\hat Q_{3j}}-2\bi{X_j}+\alpha\bi{\hat Q_3}&\lambda_{ij}\bi{X_j}\\
\lambda_{ij}\bi{X_j}&\lambda_{ij}\bi{\hat W_{3i}}
\end{bmatrix}<0, j\in{\ell_\mathcal K^i},\ j\neq i
\end{equation}

\begin{equation}\label{LMI12}
\bi{\hat Q_{3j}}-2\bi{X_j}+\bi{\hat W_{3i}}\leq0,\quad j\in{\ell_{u\mathcal K}^i},\quad j\neq i
\end{equation}

\begin{equation}\label{LMI13}
\bi{\hat Q_{3i}}-2\bi{X_i}+\bi{\hat W_{3i}}\geq0,\quad j=i
\end{equation}
where for \(j\in{\ell_\mathcal K^i},\ i\in{\ell_\mathcal K^i}\ (b=1)\)
\begin{align}\notag
\bi{\hat\theta_{1i}}=&\begin{bmatrix}
\bi{\hat\theta_{111i}}&\bi{\hat\theta_{12i}}&\bi 0&\bi 0\\
*&-(1-\mu_i)\bi{\hat Q_{2i}}&\bi 0&\bi 0\\
\bi 0&\bi 0&\bi{\hat Q_{2i}}-\bi{\hat Q_{1i}}&\bi 0\\
\bi 0&\bi 0&\bi 0&-\bi{\hat Q_{3i}}
\end{bmatrix}\\\notag
\bi{\hat\theta_{111i}}=&\left(\bi{\bar A_{11i}}\bi{X_i}-\bi{\bar A_{12i}}\bi{Y_i}\right)+\left(\bi{\bar A_{11i}}\bi{X_i}-\bi{\bar A_{12i}}\bi{Y_i}\right)^T\\
&+\bi{\hat Q_{1i}}+\bi{\hat Q_{3i}}-\sum_{j\in{\ell_\mathcal K^i}}\lambda_{ij}\bi{V_i}+\lambda_{ii}\bi{X_i},
\label{theta1i}
\end{align}
and for \(j\in{\ell_\mathcal K^i},\ i\in{\ell_{u\mathcal K}^i}\ (b=2)\)
\begin{align}\notag
\bi{\hat\theta_{2i}}=&\begin{bmatrix}
\bi{\hat\theta_{112i}}&\bi{\hat\theta_{12i}}&\bi 0&\bi 0\\
*&-(1-\mu_i)\bi{\hat Q_{2i}}&\bi 0&\bi 0\\
\bi 0&\bi 0&\bi{\hat Q_{2i}}-\bi{\hat Q_{1i}}&\bi 0\\
\bi 0&\bi 0&\bi 0&-\bi{\hat Q_{3i}}
\end{bmatrix}\\\notag
\bi{\hat\theta_{112i}}=&\left(\bi{\bar A_{11i}}\bi{X_i}-\bi{\bar A_{12i}}\bi{Y_i}\right)+\left(\bi{\bar A_{11i}}\bi{X_i}-\bi{\bar A_{12i}}\bi{Y_i}\right)^T\\
&+\bi{\hat Q_{1i}}+\bi{\hat Q_{3i}}-\sum_{j\in{\ell_\mathcal K^i}}\lambda_{ij}\bi{V_i},\\
\bi{\hat\theta_{12i}}=\ &\left(\bi{\bar A_{d11i}}\bi{X_i}-\bi{\bar A_{d12i}}\bi{Y_i}\right)\notag
\end{align}
with
\begin{equation}
\bi{\hat\phi_i}=\begin{bmatrix}
\bi{\hat M_i}\ &-\bi{\hat M_i}+\bi{\hat N_i}-\bi{\hat S_i}\ &\bi{\hat S_i}\ \ &-\bi{\hat N_i}
\end{bmatrix}
\end{equation}
\begin{align}
&\bi{\hat M_i}=\begin{bmatrix}
\bi{\hat M_{i1}}^T&\bi{\hat M_{i2}}^T&\bi{\hat M_{i3}}^T&\bi{\hat M_{i4}}^T
\end{bmatrix}^T,\\
&\bi{\hat N_i}=\begin{bmatrix}
\bi{\hat N_{i1}}^T&\bi{\hat N_{i2}}^T&\bi{\hat N_{i3}}^T&\bi{\hat N_{i4}}^T
\end{bmatrix}^T,\\
&\bi{\hat S_i}=\begin{bmatrix}
\bi{\hat S_{i1}}^T&\bi{\hat S_{i2}}^T&\bi{\hat S_{i3}}^T&\bi{\hat S_{i4}}^T
\end{bmatrix}^T,
\end{align}
{\setlength\arraycolsep{2.6pt}\begin{gather}
\bi{\hat A_{im}}=\\\begin{bmatrix}
\left(\bi{\bar A_{11i}}\bi{X_i}-\bi{\bar A_{12i}}\bi{Y_i}\right)\ &\left(\bi{\bar A_{d11i}}\bi{X_i}-\bi{\bar A_{d12i}}\bi{Y_i}\right)\ &\bi 0&\bi 0
\end{bmatrix}\notag
\end{gather}}
\begin{equation}\label{gamma}
\bi{\Gamma_i}(\bi{X_i})=\begin{bmatrix}
h_1\bi{X_i}\ &h_{21}\bi{X_i}\ &h_2\bi{X_i}\ &\bi{\hat\Gamma_i}(\bi{X_i})
\end{bmatrix},
\end{equation}
\begin{equation}\label{xi}
\bi{\Xi_i}(\bi{X_i})=\ diag\left\{h_1\bi{\hat Q_1},h_{21}\bi{\hat Q_2},h_2\bi{\hat Q_3},\bi{\hat\Xi_i}(\bi{X_i})\right\},
\end{equation}
\begin{equation*}
\begin{aligned}
\bi{\hat\Gamma_i}(\bi{X_i})=\ \bigg[&\sqrt{\lambda_{i\mathcal K_1^i}}\bi{X_i},...,\sqrt{\lambda_{i\mathcal K_{i-1}^i}}\bi{X_i},\\
&\quad\sqrt{\lambda_{i\mathcal K_{i+1}^i}}\bi{X_i},...,\sqrt{\lambda_{i\mathcal K_q^i}}\bi{X_i}\bigg],
\end{aligned}
\end{equation*}
\begin{equation*}
\bi{\hat\Xi_i}(\bi{X_i})=\ diag\left\{\bi{X_{\mathcal K_1^i}},...,\bi{X_{\mathcal K_{i-1}^i}},\bi{X_{\mathcal K_{i+1}^i}},...,\bi{X_{\mathcal K_q^i}}\right\}
\end{equation*}
in which \(\alpha\) represents the number of known elements of transition probabilities matrix for \(j\neq i\), and \(\kappa_1^i,\dots,\kappa_q^i\) are defined in (\ref{known tr}). Moreover, if the LMIs~\eqref{LMI1}-\eqref{LMI13} have a feasibility solution in terms of \(\bi{X_i}\) and \(\bi{Y_i}\), the parameter \(\bi{C_i}\)  can be computed by
\begin{equation}\label{Ci}
\bi{C_i}=\bi{Y_i}\bi{X_i}^{-1}
\end{equation}
\begin{proof}\upshape\label{proof1}

Choose the stochastic Lyapunov-Krasovskii functional candidate as follows:
\begin{equation}\label{LYP1}
\begin{aligned}
V(\bi{z_1}(t),i)=\ &V_1(\bi{z_1}(t),i)+V_2(\bi{z_1}(t),i)+V_3(\bi{z_1}(t),i)\\
&+V_4(\bi{z_1}(t),i)
\end{aligned}
\end{equation}
with
\begin{equation*}
\begin{aligned}
V_1(\bi{z_1}(t),i)=\ &\bi{z_1}^T(t)\bi{P_i}\bi{z_1}(t),\\
V_2(\bi{z_1}(t),i)=\ &\int_{t-h_1}^t{\bi{z_1}^T(s)\bi{Q_{1i}}\bi{z_1}(s)ds}\\
&+\int_{t-\tau_i(t)}^{t-h_1}{\bi{z_1}^T(s)\bi{Q_{2i}}\bi{z_1}(s)ds}\\
&+\int_{t-h_2}^t{\bi{z_1}^T(s)\bi{Q_{3i}}\bi{z_1}(s)ds},\\
V_3(\bi{z_1}(t),i)=\ &\int_{-h_1}^0{\int_{t+\theta}^t{\bi{z_1}^T(s)\bi{Q_1}\bi{z_1}(s)dsd\theta}}\\
&+\int_{-h_2}^{-h_1}{\int_{t+\theta}^t{\bi{z_1}^T(s)\bi{Q_2}\bi{z_1}(s)dsd\theta}}\\
&+\int_{-h_2}^0{\int_{t+\theta}^t{\bi{z_1}^T(s)\bi{Q_3}\bi{z_1}(s)dsd\theta}}\\
V_4(\bi{z_1}(t),i)=\ &\int_{-h_2}^0{\int_{t+\theta}^t{\bi{\dot z_1}^T(s)\bi{R_1}\bi{\dot z_1}(s)dsd\theta}}\\
&+\int_{-h_2}^{-h_1}{\int_{t+\theta}^t{\bi{\dot z_1}^T(s)\bi{R_2}\bi{\dot z_1}(s)dsd\theta}}
\end{aligned}
\end{equation*}
By using the weak infinitesimal operator of the Lyapunov function \(\mathfrak LV(\bi{z_1}(t),i)\)~\cite{bib30}, we have:
\begin{equation}\label{Lv1}
\begin{aligned}
&\mathfrak LV_1(\bi{z_1}(t),i)=\\
&\ \ \quad\quad\bi{z_1}^T(t)\left\{\bi{\tilde A_i}^T\bi{P_i}+\bi{P_i}\bi{\tilde A_i}+\sum_{j\in\ell} \lambda_{ij}\bi{P_j}\right\}\bi{z_1}(t)\\
&\quad\quad+2\bi{z_1}^T(t)\bi{P_i}\bi{\tilde A_{di}}\bi{z_1}(t-\tau_i(t))
\end{aligned}
\end{equation}
\begin{equation}\label{Lv2}
\begin{aligned}
&\mathfrak LV_2(\bi{z_1}(t),i)=\\
&\quad\ \ \quad\bi{z_1}^T(t)\left[\bi{Q_{1i}}+\bi{Q_{3i}}\right]\bi{z_1}(t)\\
&\ \quad+\bi{z_1}^T(t-h_1)\left(\bi{Q_{2i}}-\bi{Q_{1i}}\right)\bi{z_1}(t-h_1)\\
&\ \quad-\bi{z_1}^T(t-h_2)\bi{Q_{3i}}\bi{z_1}(t-h_2)\\
&\ \quad-\left(1-\dot\tau_i(t)\right)\bi{z_1}^T(t-\tau_i(t))\bi{Q_{2i}}\bi{z_1}(t-\tau_i(t))\\
&\ \quad+\sum_{j\in\ell}\int_{t-h_1}^{t}\bi{z_1}^T(s)\left(\lambda_{ij}\bi{Q_{1j}}\right)\bi{z_1}(s)ds\\
&\ \quad+\sum_{j\in\ell}\int_{t-\tau_j(t)}^{t-h_1}\bi{z_1}^T(s)\left(\lambda_{ij}\bi{Q_{2j}}\right)\bi{z_1}(s)ds\\
&\ \quad+\sum_{j\in\ell}\int_{t-h_2}^{t}\bi{z_1}^T(s)\left(\lambda_{ij}\bi{Q_{3j}}\right)\bi{z_1}(s)ds\\
&\ \ \leq\bi{z_1}^T(t)\left[\bi{Q_{1i}}+\bi{Q_{3i}}\right]\bi{z_1}(t)\\
&\ \quad+\bi{z_1}^T(t-h_1)\left(\bi{Q_{2i}}-\bi{Q_{1i}}\right)\bi{z_1}(t-h_1)\\
&\ \quad-\bi{z_1}^T(t-h_2)\bi{Q_{3i}}\bi{z_1}(t-h_2)\\
&\ \quad-(1-\mu_i)\bi{z_1}^T(t-\tau_i(t))\bi{Q_{2i}}\bi{z_1}(t-\tau_i(t))\\
&\ \quad+\sum_{j\in\ell}\int_{t-h_1}^{t}\bi{z_1}^T(s)\left(\lambda_{ij}\bi{Q_{1j}}\right)\bi{z_1}(s)ds\\
&\ \quad+\sum_{j\in\ell}\int_{t-h_2}^{t-h_1}\bi{z_1}^T(s)\left(\lambda_{ij}\bi{Q_{2j}}\right)\bi{z_1}(s)ds\\
&\ \quad+\sum_{j\in\ell}\int_{t-h_2}^{t}\bi{z_1}^T(s)\left(\lambda_{ij}\bi{Q_{3j}}\right)\bi{z_1}(s)ds\\
\end{aligned}
\end{equation}
\begin{equation}\label{Lv3}
\begin{aligned}
\mathfrak LV_3(\bi{z_1}(t),i){=}&\bi{z_1}^T(t)\left(h_1\bi{Q_1}+h_{21}\bi{Q_2}+h_2\bi{Q_3}\right)\bi{z_1}(t)\\
&-\int_{t-h_1}^t \bi{z_1}^T(s)\bi{Q_1}\bi{z_1}(s)ds\\
&-\int_{t-h_2}^{t-h1}\bi{z_1}^T(s)\bi{Q_2}\bi{z_1}(s)ds\\
&-\int_{t-h_2}^t \bi{z_1}^T(s)\bi{Q_3}\bi{z_1}(s)ds\\
\end{aligned}
\end{equation}
\begin{equation}\label{Lv4}
\begin{aligned}
\mathfrak LV_4(\bi{z_1}(t),i)=\ &\bi{\dot z_1}^T(t)\left(h_2\bi{R_1}+h_{21}\bi{R_2}\right)\bi{\dot z_1}(t)\\&-\int_{t-h_2}^t \bi{\dot z_1}^T(s)\bi{R_1}\bi{\dot z_1}(s)ds\\
&-\int_{t-h_2}^{t-h_1} \bi{\dot z_1}^T(s)\bi{R_2}\bi{\dot z_1}(s)ds
\end{aligned}
\end{equation}
where \(h_{21}=h_2-h_1\). By using Newton-Leibniz formula, for any matrices \(\bi{M_i},\bi{N_i}\) and \(\bi{S_i}\), we have:
\begin{equation}\label{Leib1}
\begin{aligned}
2\bi{\zeta_i}^T(t)\bi{M_i}\bigg[&\bi{z_1}(t)-\bi{z_1}(t-\tau_i(t))\\
&-\int_{t-\tau_i(t)}^t{\bi{\dot z_1}(s)ds}\bigg]=0
\end{aligned}
\end{equation}
\begin{equation}\label{Leib2}
\begin{aligned}
2\bi{\zeta_i}^T(t)\bi{N_i}\bigg[&\bi{z_1}(t-\tau_i(t))-\bi{z_1}(t-h_2)\\
&-\int_{t-h_2}^{t-\tau_i(t)}{\bi{\dot z_1}(s)ds}\bigg]=0
\end{aligned}
\end{equation}
\begin{equation}\label{Leib3}
\begin{aligned}
2\bi{\zeta_i}^T(t)\bi{S_i}\bigg[&\bi{z_1}(t-h_1)-\bi{z_1}(t-\tau_i(t))\\
&-\int_{t-\tau_i(t)}^{t-h_1}{\bi{\dot z_1}(s)ds}\bigg]=0
\end{aligned}
\end{equation}
where
\setlength\arraycolsep{4pt}
\begin{multline*}
{\zeta_i}(t)=\\\begin{bmatrix}
\bi{z_1}^T(t)&\bi{z_1}^T(t-\tau_i(t))&\bi{z_1}^T(t-h_1)&\bi{z_1}^T(t-h_2)
\end{bmatrix}^T
\end{multline*}
Then, by adding left sides of  (\ref{Leib1}), (\ref{Leib2}) and (\ref{Leib3}) to (\ref{Lv4}) and using Jensen's inequality~\cite{bib38} and lemma \(\ref{lem5}\), we have:

\begin{equation}
\begin{aligned}
\mathfrak LV_4(\bi{z_1}(t),i)\leq\ &\left(\bi{\tilde A_iz_1}(t)+\bi{\tilde A_{di}z_1}(t-\tau_i(t))\right)^T\cdot\\
&\quad\left(h_2\bi{R_1}+h_{21}\bi{R_2}\right)\cdot\\
&\quad\quad\left(\bi{\tilde A_iz_1}(t)+\bi{\tilde A_{di}z_1}(t-\tau_i(t))\right)\\
&+\bi{\zeta_i}^T(t)\bigg\{\tau_i(t)\bi{M_iR_1}^{-1}\bi{M_i}^T\\
&+(h_2-\tau_i(t))\bi{N_i}(\bi{R_1}+\bi{R_2})^{-1}\bi{N_i}^T\\
&+(\tau_i(t)-h_1)\bi{S_iR_2}^{-1}\bi{S_i}^T\\
&+\begin{bmatrix}
\bi{M_i}&-\bi{M_i}&\bi 0&\bi 0
\end{bmatrix}\\
&+\begin{bmatrix}
\bi{M_i}&-\bi{M_i}&\bi 0&\bi 0
\end{bmatrix}^T\\
&+\begin{bmatrix}
\bi 0&\bi{N_i}&\bi 0&-\bi{N_i}
\end{bmatrix}\\
&+\begin{bmatrix}
\bi 0&\bi{N_i}&\bi 0&-\bi{N_i}
\end{bmatrix}^T\\
&+\begin{bmatrix}
\bi 0&-\bi{S_i}&\bi{S_i}&\bi 0
\end{bmatrix}\\
&+\begin{bmatrix}
\bi 0&-\bi{S_i}&\bi{S_i}&\bi 0
\end{bmatrix}^T\bigg\}\bi{\zeta_i}(t)\\
\end{aligned}
\end{equation}
Finally,
\begin{equation}
\begin{aligned}
\mathfrak LV(\bi{z_1}(t),i)=\ &\mathfrak LV_1(\bi{z_1}(t),i)+\mathfrak LV_2(\bi{z_1}(t),i)\\
&+\mathfrak LV_3(\bi{z_1}(t),i)+\mathfrak LV_4(\bi{z_1}(t),i)
\end{aligned}
\end{equation}
If following condtions in (\ref{Lv2}) and (\ref{Lv3}) holds,
\begin{equation}\label{c1}
\sum_{j\in\ell}\lambda_{ij}\bi{Q_{1j}}\leq\ \bi{Q_1}
\end{equation}
\begin{equation}\label{c2}
\sum_{j\in\ell}\lambda_{ij}\bi{Q_{2j}}\leq\ \bi{Q_2}
\end{equation}
\begin{equation}\label{c3}
\sum_{j\in\ell}\lambda_{ij}\bi{Q_{3j}}\leq\ \bi{Q_3}
\end{equation}
Then, we can rewrite \(\mathfrak LV(\bi{z_1}(t),i)\) as follows:
\begin{equation}\label{Lvv}\setlength\arraycolsep{2pt}
\begin{aligned}
&\mathfrak LV(\bi{z_1}(t),i)\leq\\
&\ \ \quad\bi{z_1}^T(t)\bigg\{\bi{\tilde A_i}^T\bi{P_i}+\bi{P_i\tilde A_i}+\sum_{j\in\ell}\lambda_{ij}\bi{P_j}+\bi{Q_{1i}}\\
&\ \quad \quad\quad+\bi{Q_{3i}}+h_1\bi{Q_1}+h_{21}\bi{Q_2}+h_2\bi{Q_3}\bigg\}\bi{z_1}(t)\\
&\quad+\bi{z_1}^T(t-\tau_i(t))\left[-(1-\mu_i)\bi{Q_{2i}}\right]\bi{z_1}(t-\tau_i(t))\\
&\quad+\bi{z_1}^T(t-h_1)\left[\bi{Q_{2i}}-\bi{Q_{1i}}\right]\bi{z_1}(t-h_1)\\
&\quad-\bi{z_1}^T(t-h_2)\bi{Q_{3i}}\bi{z_1}(t-h_2)\\
&\quad+\bi{z_1}^T(t)\bi{P_i}\bi{\tilde A_{di}}\bi{z_1}(t-\tau_i(t))\\
&\quad+\bi{z_1}^T(t-\tau_i(t))\bi{\tilde A_{di}}^T\bi{P_i}\bi{z_1}(t)\\
&\quad+\left(\bi{\tilde A_iz_1}(t)+\bi{\tilde A_{di}z_1}(t-\tau_i(t))\right)^T\cdot\\
&\quad\quad\left(h_2\bi{R_1}+h_{21}\bi{R_2}\right)\left(\bi{\tilde A_iz_1}(t)+\bi{\tilde A_{di}z_1}(t-\tau_i(t))\right)\\
&\quad+\bi{\zeta_i}^T(t)\bigg\{\tau_i(t)\bi{M_iR_1}^{-1}\bi{M_i}^T\\
&\quad\quad\quad+(h_2-\tau_i(t))\bi{N_i}(\bi{R_1}+\bi{R_2})^{-1}\bi{N_i}^T\\
&\quad\quad\quad+(\tau_i(t)-h_1)\bi{S_i}\bi{R_2}^{-1}\bi{S_i}^T\\
&\quad\quad\quad+\begin{bmatrix}
\bi{M_i}\ &\bi{-M_i}+\bi{N_i}-\bi{S_i}\ \ &\bi{S_i}\ &-\bi{N_i}
\end{bmatrix}\\
&\quad\quad\quad+\begin{bmatrix}
\bi{M_i}\ &\bi{-M_i}+\bi{N_i}-\bi{S_i}\ \ &\bi{S_i}\ &-\bi{N_i}
\end{bmatrix}^T\bigg\} \bi{\zeta_i}(t)\\
\end{aligned}
\end{equation}
Now, by Substituting \(\bi{\tilde A_i}\) and \(\bi{\tilde A_{di}}\) from (\ref{sl}) into (\ref{Lvv}), we have:
\begin{equation}
\begin{aligned}
&\mathfrak LV(\bi{z_1}(t),i)\leq\\
&\quad\quad\bi{\zeta_i}^T(t)\bigg\{\bi{\theta_i}+\bi{\phi_i}+\bi{\phi_i}^T+\bi{A_{im}}^Th_2\bi{R_1A_{im}}\\
&\ \quad\quad\quad+\bi{A_{im}}^Th_{21}\bi{R_2}\bi{A_{im}}+\tau_i(t)\bi{M_iR_1}^{-1}\bi{M_i}^T\\
&\ \quad\quad\quad+(h_2-\tau_i(t))\bi{N_i}\left(\bi{R_1}+\bi{R_2}\right)^{-1}\bi{N_i}^T\\
&\ \quad\quad\quad+(\tau_i(t)-h_1)\bi{S_iR_2}^{-1}\bi{S_i}^T\bigg\}\bi{\zeta_i}(t)
\end{aligned}
\end{equation}
with
\begin{equation*}
\bi{\theta_i}=\begin{bmatrix}
\bi{\theta_{11i}}&\bi{\theta_{12i}}&\bi 0&\bi 0\\
*&-(1-\mu_i)\bi{Q_{2i}}&\bi 0&\bi 0\\
\bi 0&\bi 0&\bi{Q_{2i}}-\bi{Q_{1i}}&\bi 0\\
\bi 0&\bi 0&\bi 0&-\bi{Q_{3i}}
\end{bmatrix},
\end{equation*}
\begin{equation*}
\begin{aligned}
\bi{\theta_{11i}}=\ &\left(\bi{\bar A_{11i}}-\bi{\bar A_{12i}}\bi{C_i}\right)^T\bi{P_i}+\bi{P_i}\left(\bi{\bar A_{11i}}-\bi{\bar A_{12i}}\bi{C_i}\right)\\
&+\sum_{j\in\ell}\lambda_{ij}\bi{P_j}+\bi{Q_{1i}}+\bi{Q_{3i}}+h_1\bi{Q_1}\\
&+h_{21}\bi{Q_2}+h_2\bi{Q_3},
\end{aligned}
\end{equation*}
\begin{equation*}
\begin{aligned}
&\bi{\theta_{12i}}=\bi{P_i}\left(\bi{\bar A_{d11i}}-\bi{\bar A_{d12i}}\bi{C_i}\right),\\
&\bi{\phi_i}=\begin{bmatrix}
\bi{M_i}\ &-\bi{M_i}+\bi{N_i}-\bi{S_i}\ \ &\bi{S_i}\ &-\bi{N_i}
\end{bmatrix},\\
\end{aligned}
\end{equation*}
\begin{multline*}
\bi{A_{im}}=\begin{bmatrix}
\bi{\tilde A_i}\ &\bi{\tilde A_{di}}\ &\bi 0\ &\bi 0
\end{bmatrix}\\
=\begin{bmatrix}
\left(\bi{\bar A_{11i}}-\bi{\bar A_{12i}}\bi{C_i}\right)\ &\left(\bi{\bar A_{d11i}}-\bi{\bar A_{d12i}}\bi{C_i}\right)\ &\bi 0\ &\bi 0
\end{bmatrix}
\end{multline*}
Obviously, from lemma \ref{lem12} and Schur complement, we can see \(\mathfrak LV(\bi{z_1}(t),i)<0\) if following conditions hold:
\begin{multline}
\bi{\Delta_{1i}}=\left[\begin{array}{ccc}
\bi{\theta_i}+\bi{\phi_i}+\bi{\phi_i}^T&h_2\bi{A_{im}}^T&h_{21}\bi{A_{im}}^T\\
*&-h_2\bi{R_1}^{-1}&\bi 0\\
*&\bi 0&-h_{21}\bi{R_2}^{-1}\\
*&*&*\\
*&*&*
\end{array}\right.\\
\left.\begin{array}{cc}
h_1\bi{M_i}&h_{21}\bi{N_i}\\
\bi 0&\bi 0\\
\bi 0&\bi 0\\
-h_1\bi{R_1}&\bi 0\\
\bi 0&-h_{21}(\bi{R_1}+\bi{R_2})
\end{array}\right]
<0
\end{multline}
\begin{multline}
\bi{\Delta_{2i}}=\left[\begin{array}{ccc}
\bi{\theta_i}+\bi{\phi_i}+\bi{\phi_i}^T&h_2\bi{A_{im}}^T&h_{21}\bi{A_{im}}^T\\
*&-h_2\bi{R_1}^{-1}&\bi 0\\
*&\bi 0&-h_{21}\bi{R_2}^{-1}\\
*&*&*\\
*&*&*
\end{array}\right.\\
\left.\begin{array}{cc}
h_2\bi{M_i}&h_{21}\bi{S_i}\\
\bi 0&\bi 0\\
\bi 0&\bi 0\\
-h_2\bi{R_1}&\bi 0\\
\bi 0&-h_{21}\bi{R_2}
\end{array}\right]
<0
\end{multline}
By pre- and post-multiplying both sides of \(\bi{\Delta_{1i}}\) and \(\bi{\Delta_{2i}}\) with \(\bi{\hat X_i}=diag\left\{\bi{\tilde X_i},\bi{I},\bi{I},\bi{X_i},\bi{X_i}\right\}\), respectively, where \(\bi{\tilde X_i}=diag\left\{\bi{X_i},\bi{X_i},\bi{X_i},\bi{X_i}\right\}\) and \(\bi{X_i}=\bi{P_i}^{-1}\), we can define:
\begin{equation*}
\begin{aligned}
&\bi{Y_i}=\bi{C_i}\bi{X_i},\quad\bi{\hat R_1}=\bi{R_1}^{-1},\quad \bi{\hat R_2}=\bi{R_2}^{-1},\\
&\bi{\hat Q_{1i}}=\bi{X_iQ_{1i}X_i},\ \bi{\hat Q_{2i}}=\bi{X_iQ_{2i}X_i},\ \bi{\hat Q_{3i}}=\bi{X_iQ_{3i}X_i},\\
&\bi{\hat M_i}=\bi{X_iM_iX_i},\quad \bi{\hat N_i}=\bi{X_iN_iX_i},\quad \bi{\hat S_i}=\bi{X_iS_iX_i},\\
&\bi{\hat Q_1}=\bi{Q_1}^{-1},\quad \bi{\hat Q_2}=\bi{Q_2}^{-1},\quad \bi{\hat Q_3}=\bi{Q_3}^{-1},\\
&\bi{\hat\phi_i}=\begin{bmatrix}
\bi{\hat M_i}\ &-\bi{\hat M_i}+\bi{\hat N_i}-\bi{\hat S_i}\ \ &\bi{\hat S_i}\ &-\bi{\hat N_i}
\end{bmatrix},
\end{aligned}
\end{equation*}
\begin{align*}
\bi{\hat A_{im}}&=\bi{A_{im}X_i}\\
&=\big[\begin{array}{l}\left(\bi{\bar A_{11i}X_i}-\bi{\bar A_{12i}Y_i}\right)\end{array}\\
&\qquad\qquad\begin{array}{lll}\left(\bi{\bar A_{d11i}X_i}-\bi{\bar A_{d12i}Y_i}\right)\ &\bi 0\ &\bi 0\end{array}\big]
\end{align*}
Then, we have
\begin{multline}
\bi{\tilde\Delta_{1i}}=\left[\begin{array}{ccc}
\bi{\tilde\theta_i}+\bi{\hat\phi_i}+\bi{\hat\phi_i}^T&h_2\bi{\hat A_{im}}^T&h_{21}\bi{\hat A_{im}}^T\\
*&-h_2\bi{\hat R_1}&\bi 0\\
*&\bi 0&-h_{21}\bi{\hat R_2}\\
*&*&*\\
*&*&*
\end{array}\right.\\
\left.\begin{array}{cc}
h_1\bi{\hat M_i}&h_{21}\bi{\hat N_i}\\
\bi 0&\bi 0\\
\bi 0&\bi 0\\
-h_1\bi{X_iR_1X_i}&\bi 0\\
\bi 0&-h_{21}\bi{X_i}(\bi{R_1}+\bi{R_2})\bi{X_i}
\end{array}\right]
<0
\end{multline}
\begin{multline}
\bi{\tilde\Delta_{2i}}=\left[\begin{array}{ccc}
\bi{\tilde\theta_i}+\bi{\hat\phi_i}+\bi{\hat\phi_i}^T&h_2\bi{\hat A_{im}}^T&h_{21}\bi{\hat A_{im}}^T\\
*&-h_2\bi{\hat R_1}&\bi 0\\
*&\bi 0&-h_{21}\bi{\hat R_2}\\
*&*&*\\
*&*&*
\end{array}\right.\\
\left.\begin{array}{cc}
h_2\bi{\hat M_i}&h_{21}\bi{\hat S_i}\\
\bi 0&\bi 0\\
\bi 0&\bi 0\\
-h_2\bi{X_iR_1X_i}&\bi 0\\
\bi 0&-h_{21}\bi{X_iR_2X_i}
\end{array}\right]
<0
\end{multline}

with
\begin{equation}
\bi{\tilde\theta_i}=\begin{bmatrix}
\bi{\tilde\theta_{11i}}&\bi{\hat\theta_{12i}}&\bi 0&\bi 0\\
*&-(1-\mu_i)\bi{\hat Q_{2i}}&\bi 0&\bi 0\\
\bi 0&\bi 0&\bi{\hat Q_{2i}}-\bi{\hat Q_{1i}}&\bi 0\\
\bi 0&\bi 0&\bi 0&-\bi{\hat Q_{3i}}
\end{bmatrix}
\end{equation}

\begin{equation*}
\begin{aligned}
\bi{\tilde\theta_{11i}}=\ &\left(\bi{\bar A_{11i}X_i}-\bi{\bar A_{12i}Y_i}\right)+\left(\bi{\bar A_{11i}X_i}-\bi{\bar A_{12i}Y_i}\right)^T\\
&+\bi{X_i}\sum_{j\in\ell}\lambda_{ij}\bi{X_j}^{-1}\bi{X_i}+\bi{\hat Q_{1i}}+\bi{\hat Q_{3i}}\\
&+h_1\bi{X_iQ_1X_i}+h_{21}\bi{X_iQ_2X_i}+h_2\bi{X_iQ_3X_i},\\
\bi{\hat\theta_{12i}}=\ &\left(\bi{\bar A_{d11i}X_i}-\bi{\bar A_{d12i}Y_i}\right)
\end{aligned}
\end{equation*}
Since \(\sum_{j\in{\ell}}\lambda_{ij}=0\), we can rewrite \(\bi{\tilde\theta_{11i}}\) as follows:
\begin{align}\notag
\bi{\tilde\theta_{11i}}=\ &\left(\bi{\bar A_{11i}X_i}-\bi{\bar A_{12i}Y_i}\right)+\left(\bi{\bar A_{11i}X_i}-\bi{\bar A_{12i}Y_i}\right)^T\\\notag
&+\bi{\hat Q_{1i}}+\bi{\hat Q_{3i}}+\sum_{j\in{\ell}}\lambda_{ij}\left(\bi{X_iX_j}^{-1}\bi{X_i}\right)\\
&-\sum_{j\in{\ell}}\lambda_{ij}\bi{V_i}+h_1\bi{X_iQ_1X_i}\\\notag
&+h_{21}\bi{X_iQ_2X_i}+h_2\bi{X_iQ_3X_i},
\end{align}
where \(\bi{V_i}=\bi{V_i}^T\) are free-connection weighting matrices. In order to solve the problem of MJSs with partly unknown transition probability rates, we separate the known and unknown elements of transition probabilities matrix by using (\ref{lambda}):
\begin{align}\notag
\bi{\tilde\theta_{11i}}=\ &\left(\bi{\bar A_{11i}X_i}-\bi{\bar A_{12i}Y_i}\right)+\left(\bi{\bar A_{11i}X_i}-\bi{\bar A_{12i}Y_i}\right)^T\\\notag
&+\bi{\hat Q_{1i}}+\bi{\hat Q_{3i}}+\sum_{j\in{\ell_\mathcal K^i}}\lambda_{ij}\left(\bi{X_iX_j}^{-1}\bi{X_i}-\bi{V_i}\right)\\\label{theta11i}
&+\sum_{j\in{\ell_{u\mathcal K}^i}}\lambda_{ij}\left(\bi{X_iX_j}^{-1}\bi{X_i}-\bi{V_i}\right)\\\notag
&+h_1\bi{X_iQ_1X_i}+h_{21}\bi{X_iQ_2X_i}+h_2\bi{X_iQ_3X_i},
\end{align}
In fact, with usage of free-connection weighting matrices instead of fixed ones, less-conservative stability criterion for MJSs with partly unknown transition probability rates is obtained. 

Note that, for \(\bi{Q>0}\), the following matrix inequality always holds:
\begin{equation}
\begin{aligned}
\left(\bi{X_i}-\bi{Q}^{-1}\right)\bi{Q}&\left(\bi{X_i}-\bi{Q}^{-1}\right)=\\
&\quad\quad\quad\bi{X_iQX_i}-2\bi{X_i}+\bi{Q}^{-1}\geq 0
\end{aligned}
\end{equation}
and so we have:
\begin{equation}\label{inequality}
-\bi{X_iQX_i}\leq -2\bi{X_i}+\bi{Q}^{-1}
\end{equation}
which is a useful tool for the development of our results.
Now, by using Schur complement, it is obvious that \(\bi{\hat\Delta_{1i}}\)~\eqref{LMI1} and \(\bi{\hat\Delta_{2i}}\)~\eqref{LMI2} in Theorem~\ref{thrm1}, are obtained from \(\bi{\tilde\Delta_{1i}}\) and \(\bi{\tilde\Delta_{2i}}\), for each \(i\in{\ell_\mathcal K^i}\ (b=1)\), and \(i\in{\ell_{u\mathcal K}^i}\ (b=2)\) in the LMI framework. In which, \(\hat\theta_{bi}\) for each \(i\in{\ell_\mathcal K^i}\ (b=1)\), and \(i\in{\ell_{u\mathcal K}^i}\ (b=2)\) are expressed in Theorem~\ref{thrm1}.

Besides, we consider unknown elements of probabilities matrix in (\ref{theta11i}) as follows: 
\begin{equation}\label{unknown}
\sum_{j\in{\ell_{u\mathcal K}^i}}\lambda_{ij}\left(\bi{X_iX_j}^{-1}\bi{X_i}-\bi{V_i}\right)\leq0
\end{equation}
Since \(\lambda_{ij}\geq 0\ (i,j\in \ell,j\neq i)\) and \(\lambda_{ii}<0\ (\lambda_{ii}=-\sum_{j=1,j\neq i}^N{\lambda_{ij}})\), it is straightforward that (\ref{unknown}) holds if the sets of LMIs~\eqref{LMI3} and~\eqref{LMI4} in Theorem~\ref{thrm1} satisfy for \(i\in{\ell_\mathcal K^i}\) and \(i\in{\ell_{u\mathcal K}^i}\), respectively.

Furthermore, in order to consider unknown elements of probabilities matrix in (\ref{c1})--(\ref{c3}), by using free connection weighting matrices \(\bi{W_{ri}}=\bi{W_{ri}}^T\) with \(r=1,2,3\), and separating the known and unknown elements of transition probabilities matrix by (\ref{lambda}), we have:
\begin{equation}\label{e1}
\sum_{j\in{\ell_\mathcal K^i}}\!\lambda_{ij}(\bi{Q_{1j}}-\bi{W_{1i}})+\!\sum_{j\in{\ell_{u\mathcal K}^i}}\!\lambda_{ij}(\bi{Q_{1j}}-\bi{W_{1i}})\leq \bi{Q_1}
\end{equation}
\begin{equation}\label{e2}
\sum_{j\in{\ell_\mathcal K^i}}\!\lambda_{ij}(\bi{Q_{2j}}-\bi{W_{2i}})+\!\sum_{j\in{\ell_{u\mathcal K}^i}}\!\lambda_{ij}\left(\bi{Q_{2j}}-\bi{W_{2i}}\right)\leq \bi{Q_2}
\end{equation}
\begin{equation}\label{e3}
\sum_{j\in{\ell_\mathcal K^i}}\lambda_{ij}\!\left(\bi{Q_{3j}}-\bi{W_{3i}}\right)+\!\sum_{j\in{\ell_{u\mathcal K}^i}}\!\lambda_{ij}\left(\bi{Q_{3j}}-\bi{W_{3i}}\right)\leq \bi{Q_3}
\end{equation}
At first we consider (\ref{e1}) and rewrite it as follows:
\begin{align}\notag
&\lambda_{ii}\left(\bi{Q_{1i}}-\bi{W_{1i}}\right)\\
\notag&+\sum_{j\in{\ell_\mathcal K^i},j\neq i}\left(\lambda_{ij}\left(\bi{Q_{1j}}-\bi{W_{1i}}\right)-\frac{1}{\alpha}\bi{Q_1}\right)\\
&\ +\sum_{j\in{\ell_{u\mathcal K}^i},j\neq i}\lambda_{ij}\left(\bi{Q_{1j}}-\bi{W_{1i}}\right)\leq0
\label{e11}
\end{align}
where \(\alpha\) is the number of known elements of transition probabilities matrix for \(j\neq i\). Then, it is straightforward that (\ref{e11}) holds if the following set of LMIs satisfies 
\begin{equation}\label{v1}
\sum_{j\in{\ell_\mathcal K^i},j\neq i}\left(\lambda_{ij}\left(\bi{Q_{1j}}-\bi{W_{1i}}\right)-\frac {1}{\alpha} \bi{Q_1}\right)<0
\end{equation}
and
\begin{equation}\label{v2}
\sum_{j\in{\ell_{u\mathcal K}^i},j\neq i}\lambda_{ij}\left(\bi{Q_{1j}}-\bi{W_{1i}}\right)\leq0
\end{equation}
and
\begin{equation}\label{v3}
\lambda_{ii}\left(\bi{Q_{1i}}-\bi{W_{1i}}\right)\leq0
\end{equation}
In (\ref{v1}), following LMI holds for each \(j\in{\ell_\mathcal K^i},\ j\neq i\):
\begin{equation}\label{v11}
\lambda_{ij}\left(\bi{Q_{1j}}-\bi{W_{1i}}\right)-\frac{1}{\alpha}\bi{Q_1}<0
\end{equation}
Pre- and post-multiplying both sides of (\ref{v11}) with \(\bi{X_j}\) and using Schur complement and the inequality in~\eqref{inequality} yields:
\begin{equation}
\begin{bmatrix}
\lambda_{ij}\bi{\hat Q_{1j}}-2\bi{X_j}+\alpha\bi{\hat Q_1}&\lambda_{ij}\bi{X_j}\\
\lambda_{ij}\bi{X_j}&\lambda_{ij}\bi{\hat W_{1i}}
\end{bmatrix}<0
\end{equation}
In (\ref{v2}), the following LMI holds for each \(j\in{\ell_{u\mathcal K}^i},\ j\neq i\):
\begin{equation}\label{v22}
\left(\bi{Q_{1j}}-\bi{W_{1i}}\right)\leq0
\end{equation}
By pre- and post-multiplying both sides of (\ref{v22}) with \(\bi{X_j}\), and using the inequality in~\eqref{inequality}, we have:
\begin{equation}
\bi{\hat Q_{1j}}-2\bi{X_j}+\bi{\hat W_{1i}}\leq0
\end{equation}
Since \(\lambda_{ii}<0\), in (\ref{v3}), the following LMI holds for \(j=i\):
\begin{equation}\label{v33}
\bi{Q_{1i}}-\bi{W_{1i}}\geq 0
\end{equation}
Pre- and post-multiplying both sides of (\ref{v33}) with \(\bi{X_i}\), and considering the inequality in~\eqref{inequality} yields:
\begin{equation}
\bi{\hat Q_{1i}}-2\bi{X_i}+\bi{\hat W_{1i}}\geq0
\end{equation}
where \(\bi{\hat Q_1}=\bi{Q_1}^{-1}, \bi{\hat W_{1i}}=\bi{W_{1i}}^{-1}, \bi{\hat Q_{1i}}=\bi{X_i Q_i X_i}\).

The same above calculation is applied to (\ref{e2}) and (\ref{e3}), respectively.
Therefore, by considering (\ref{theta1i})-(\ref{xi}), if the set of LMIs (\ref{LMI1})-(\ref{LMI13}) is satisfied; i.e. \(\bi{C_i}=\bi{Y_iX_i}^{-1}\), then \(\mathfrak LV(\bi{z_1}(t),i)<0\) and stochastic stability of sliding mode dynamics (\ref{sl}) is verified. This concludes the proof.
\end{proof}
\end{theorem}
\begin{remark}\label{remark2}
In the Theorem~\ref{thrm1}, the sets of LMIs (\ref{LMI1})-(\ref{LMI13}) guaranteeing stochastic stability of sliding mode dynamics~\eqref{sl} are delay-dependent. In other words, the obtained conditions include the information on the size of delay and its derivative; i.e. \(h_1\), \(h_2\) and \(\mu_i\). Note that by using these bounds, less-conservative stability criteria is obtained. To the best of our knowledge, in most of the studies such as~\cite{bib41,bib42,bib43,bib44,bib47}, derivative of mode-dependent time-varying delay is assumed to be less than one. In this paper, we use new techniques to remove the derivative restriction and to obtain less-conservative criteria. To reach this purpose, we construct a novel stochastic Lyapunov-Krasovskii functional candidate including the information of delay size. Therefore, in our design method, the derivative of the mode-dependent time-varying delay may be larger than one; i.e. in this paper, the mode-dependent time-varying delay is more general.
\end{remark}
\begin{remark}\label{remark3}
In the proof process, we encounter some nonlinear terms in the form of~\(\bi{X_iQX_i}\) with \(\bi{Q}>0\) that makes it hard to express stochastic stability criterion in terms of LMIs. In order to solve this problem, some assumptions have been presented by other researchers such as in~\cite{bib46,bib45} which may not be satisfied. In fact, the authors in~\cite{bib46,bib45}  defined new variables such as \(\bi{R_i}=\bi{X_iQX_i}\) or \(\bi{R}=\bi{X_iQX_i}\), where \(\bi{X_i}\) and \(\bi{R_i}\) or \(\bi R\) are design parameters. However, these equalities are impossible to fulfill for each~\(i\in\ell\) because of the mode-independent matrix~\(\bi Q\). In detail, finding a constant~\(\bi Q\) for all obtained design parameters \(\bi{X_i}\) and \(\bi{R_i}\) or \(\bi R\) such that~\(\bi Q=\bi{X_i}^{-T}\bi{R_iX_i}^{-1}\) or~\(\bi Q=\bi{X_i}^{-T}\bi{RX_i}^{-1}\), for each~\(i\in\ell\) are generally impossible~\cite{comment}. In this paper, we use a new approach to deal with this constraint and finally derive stochastic stability conditions in terms of LMIs.
\end{remark}
In the following theorem, a sliding mode controller \(\bi u(t)\) is synthesized to guarantee the reachability of sliding surface \(\bi s(t)=0\) for each \(i\) for  Markovian jump systems with mode-dependent time-varying delays and partly unknown transition probability rates.

\begin{theorem}\label{thrm2}
Consider the  Markovian jump system~\eqref{regular} with mode-dependent time-varying delays~\(\tau_i(t)\) and partly unknown transition probabilities~\eqref{lambda}. Suppose that the linear sliding surface is given by~\eqref{sliding surface} and \(\bi{C_i}\) is obtained in Theorem~\ref{thrm1}, and there exist matrices \(\bi{\Omega_i}>0\)   and \(\bi{\hat V_i}=\bi{\hat V_i}^T\)  such that the following sets of LMIs hold for each \(i\in\ell\):
\begin{equation}\label{LMI-U1}
\bi{\Omega_j}-\bi{\hat V_i}\leq0,\quad j\in{\ell_{u\mathcal K}^i,j\neq i}
\end{equation}
\begin{equation}\label{LMI-U2}
\bi{\Omega_j}-\bi{\hat V_i}\geq0,\quad j\in{\ell_{u\mathcal K}^i,j=i}
\end{equation}
Then, the state trajectories of system~\eqref{regular} can be reached the sliding surface \(\bi s(t)=0\) in the finite time by the following SMC law: 
\begin{align}\notag
\bi u(t)=&-\left(\bi{C_{2i}}\bi{B_{2i}}\right)^{-1}\bigg\{\left[\begin{array}{ll}
\bi{C_{1i}}&\bi{C_{2i}}
\end{array}\right]\cdot\\\notag
&\quad\quad\quad\quad\quad\quad\quad\left(\bi{\bar A_iz}(t)+\bi{\bar A_{di}z}(t-\tau_i(t))\right)\bigg\}\\[1mm]
&-\left(\epsilon_i+f_i\right)\mathrm{sign}\left(\bi{B_{2i}}^T\bi{C_{2i}}^T\bi{\Omega_is}(t)\right)\label{u}\\
&-\frac12\left(\bi{\Omega_i}\bi{C_{2i}}\bi{B_{2i}}\right)^{-1}\sum_{j\in{\ell_\mathcal K^i}}\lambda_{ij}\left(\bi{\Omega_j}-\bi{\hat V_i}\right)s(t)
\notag
\end{align}
where  \(\epsilon_i>0\) is a given small constant.
\begin{proof}\upshape
Choose the appropriate mode-dependent Lyapunov function candidate as
\begin{equation}
V(\bi z,t,i)=\frac12 \bi{s}^T(t)\bi{\Omega_is}(t)
\end{equation}
According to~\eqref{regular} and~\eqref{sliding surface}, we have
\begin{equation}\label{s-dot}
\begin{aligned}
\bi{\dot s}(t)=\ &\left[\begin{array}{ll}
\bi{C_{1i}}&\bi{C_{2i}}
\end{array}\right]\bigg\{\bi{\bar A_iz}(t)+\bi{\bar A_{di}z}(t-\tau_i(t))\\
&\quad\quad\quad\quad\quad\ + \left[\begin{array}{l}
\bi 0\\
\bi{B_{2i}}
\end{array}\right][\bi{u}(t)+\bi{F_iw}(t)]\bigg\}
\end{aligned}
\end{equation}
Applying the weak infinitesimal operator of the Lyapunov function \(\mathfrak LV(\bi z,t,i)\) and using~\eqref{s-dot}, yields
\begin{equation}\label{LV1}
\begin{aligned}
&\mathfrak LV(\bi z,t,i)=\bi s^T(t)\bi{\Omega_i}\\ 
&\quad\quad\times\bigg\{\left[\begin{array}{ll}
\bi{C_{1i}}&\bi{C_{2i}}
\end{array}\right]\left(\bi{\bar A_iz}(t)+\bi{\bar A_{di}z}(t-\tau_i(t)\right)\bigg\}\\
&\quad+\bi s^T(t)\bi{\Omega_iC_{2i}B_{2i}}[\bi u(t)+\bi{F_iw}(t)]\\
&\quad+\frac12\sum_{j\in{\ell}}\lambda_{ij}\bi s^T(t)\bi{\Omega_js}(t)
\end{aligned}
\end{equation}
From \(\sum_{j\in{\ell}}\lambda_{ij}=0\), it follows that following equation holds for arbitrary matrices \(\bi{\hat V_i}=\bi{\hat V_i}^T\)
\begin{equation}\label{eq=zero}
-\frac12\sum_{j\in{\ell}}\lambda_{ij}\bi s^T(t)\bi{\hat V_is}(t)=0,\quad i\in\ell
\end{equation}
Adding the left side of~\eqref{eq=zero} into~\eqref{LV1} and separating the known and unknown elements of transition probabilities matrix by (\ref{lambda}), yields:
\begin{equation}\label{LV2}
\begin{aligned}
\mathfrak LV(\bi z,t,i)=&\ \bi s^T(t)\bi{\Omega_i}\bigg\{\left[\begin{array}{ll}
\bi{C_{1i}}&\bi{C_{2i}}
\end{array}\right]\cdot\\
&\quad\quad\left(\bi{\bar A_iz}(t)+\bi{\bar A_{di}z}(t-\tau_i(t))\right)\bigg\}\\
&+\bi s^T(t)\bi{\Omega_iC_{2i}B_{2i}}[\bi u(t)+\bi{F_iw}(t)]\\
&+\frac12\sum_{j\in{\ell_\mathcal K^i}}\lambda_{ij}\left(\bi s^T(t)\left(\bi{\Omega_j}-\bi{\hat V_i}\right)\bi s(t)\right)\\
&+\frac12\sum_{j\in{\ell_{u\mathcal K}^i}}\lambda_{ij}\left(\bi s^T(t)\left(\bi{\Omega_j}-\bi{\hat V_i}\right)\bi s(t)\right)
\end{aligned}
\end{equation}
By substituting SMC law~\eqref{u} into~\eqref{LV2}, we have
\begin{equation}\label{eq82}
\begin{aligned}
\mathfrak LV(\bi z,t,i)=&\ \bi s^T(t)\bi{\Omega_iC_{2i}B_{2i}}\bigg\{-\left(\epsilon_i+f_i\right)\cdot\\
&\quad\quad\mathrm{sign}\left(\bi{B_{2i}}^T\bi{C_{2i}}^T\bi{\Omega_is}(t)\right)+\bi{F_iw}(t)\bigg\}\\
&+\frac12\sum_{j\in{\ell_{u\mathcal K}^i}}\lambda_{ij}\left(\bi s^T(t)\left(\bi{\Omega_j}-\bi{\hat V_i}\right)\bi s(t)\right)
\end{aligned}
\end{equation}
Note that if the sets of LMIs~\eqref{LMI-U1} and~\eqref{LMI-U2} hold for  \(i\in{\ell_{\mathcal K}^i}\) and \(i\in{\ell_{u\mathcal K}^i}\), then the following inequalities hold.
\begin{equation}\label{LMIU1-U2}
\frac12\sum_{j\in{\ell_{u\mathcal K}^i}}\lambda_{ij}\left(\bi s^T(t)\left(\bi{\Omega_j}-\bi{\hat V_i}\right)\bi s(t)\right)<0
\end{equation}
Thus, from~\eqref{LMI-U1} and~\eqref{LMI-U2}, we have
\begin{equation}
\begin{aligned}
\mathfrak LV(\bi z,t,i)&\leq-\left(\epsilon_i+f_i\right)\left\Vert\bi{ B_{2i}}^T\bi{C_{2i}}^T\bi{\Omega_is}(t)\right\Vert\\
&\ +f_i\left\Vert \bi{B_{2i}}^T\bi{C_{2i}}^T\bi{\Omega_is}(t)\right\Vert\\&\leq-\epsilon_i\left\Vert \bi{B_{2i}}^T\bi{C_{2i}}^T\bi{\Omega_is}(t)\right\Vert<0
\end{aligned}
\end{equation}
note that
\begin{equation}
\begin{aligned}
\left\Vert \bi{\Omega_is}(t)\right\Vert^2=& \left(\bi{\Omega_i}^{\frac12}\bi s^T(t)\right)^T\bi{\Omega_i}\left(\bi{\Omega_i}^{\frac12}\bi s^T(t)\right)\\
&\ \geq \lambda_{min}\left(\bi{\Omega_i}\right)\left\Vert \bi{\Omega_i}^\frac12\bi{s}^T(t)\right\Vert^2
\end{aligned}
\end{equation}
and so we have
\begin{equation}\label{tr}
\begin{aligned}
&\mathfrak LV(\bi z,t,i)\leq -\varrho_iV^\frac12(\bi z,t,i)\\
&\varrho_i =\sqrt 2\epsilon_i\min_{i\in\ell}\left(\lambda_{min}(\bi{\Omega_i})\right)^\frac12\times\\
&\quad\quad\ \quad\quad\quad\min_{i\in\ell} \left\{\left(\lambda_{min}(\bi{C_{2i}B_{2i}B_{2i}}^T\bi{C_{2i}}^T)\right)^\frac12\right\}
\end{aligned}
\end{equation}
where \(\varrho_i>0\). Using Dynkin's formula~\cite{bib48}, this yields
\begin{equation}
2E\left[V(\bi z(t),r(t)\right]^\frac12 \leq -\varrho_i t+2V^\frac12(\bi{z}(0),r(0))
\end{equation}
Thus, there exists an instant \(t^\star={2V^\frac12(\bi{z_0},r_0)}/{\varrho_i}\) such that \(V(\bi z,t,i)=0\), and consequently \(\bi s(t)=0\), for \(t>t^\star\). Therefore, by applying the sliding control law~\eqref{u}, the state trajectories of closed-loop system can enter the desired sliding surface~\eqref{sliding surface} in finite time. This completes the proof.
\end{proof}
\end{theorem}

{\begin{remark}\label{remark4}
Note that in some practical situations, the time delay functions \(\tau_i(t), i=1,2,...,N\) are not explicitly known a priori, and consequently, the desired delay states \(z(t-\tau_i(t))\) cannot be employed in the control law~\eqref{u} in these cases. To cope with this kind of practical problems, this paper also proposes another sliding mode controller in Theorem~\ref{thrm3}  for considered systems with unknown mode-dependent time-varying delays~\(\tau_i(t)\).
\end{remark}}
Before proceeding, we give the following assumption:
\begin{assumption}\label{assump}
According to the Razumikhin Theorem~\cite{bib777}, there exists a constant \(r>0\) such that the following inequality holds:
\begin{equation}\label{raz}
\Vert\bi z(t+\theta)\Vert\leq r\Vert\bi z(t)\Vert,\quad \ \theta\in\left[-d,0\right]
\end{equation}
In~\eqref{raz}, \(r\) is an unknown constant which should be first estimated by designing an adaptive law. If \(r(t)\) represents the estimate of \(r\), we have the following estimation error:
\begin{equation}
\tilde r(t)=r(t)-r
\end{equation}
\end{assumption}
Now we can present the following theorem to obtain an adaptive sliding mode control law for system~\eqref{regular} with unknown mode-dependent time-varying delays.
\begin{theorem}\label{thrm3}
Consider the  Markovian jump system~\eqref{regular} with {unknown} mode-dependent time-varying delays~\(\tau_i(t)\) and partly unknown transition probabilities~\eqref{lambda}. Suppose that the linear sliding surface is given by~\eqref{sliding surface} and \(\bi{C_i}\) is obtained in Theorem~\ref{thrm1}, and there exist matrices \(\bi{\Omega_i}>0\)   and \(\bi{\hat H_i}=\bi{\hat H_i}^T\)  such that the following sets of LMIs hold for each \(i\in\ell\):
\begin{equation}\label{LMI-U11}
\bi{\Omega_j}-\bi{\hat H_i}\leq0,\quad j\in{\ell_{u\mathcal K}^i,j\neq i}
\end{equation}
\begin{equation}\label{LMI-U22}
\bi{\Omega_j}-\bi{\hat H_i}\geq0,\quad j\in{\ell_{u\mathcal K}^i,j=i}
\end{equation}
Then, the state trajectories of system~\eqref{regular} can reach the sliding surface \(\bi s(t)=0\) in finite time {by the SMC law~\eqref{u2} and the adaptive law~\eqref{rdot}}.
\begin{figure*}[!ht]
\begin{equation}
\begin{aligned}\label{u2}
\bi u(t)=&-\left(\frac1{\sqrt{\lambda_{min}\left(\bi{C_{2i}}\bi{B_{2i}}\right)^T\left(\bi{C_{2i}}\bi{B_{2i}}\right)}}\right)
\times\\&\hspace{2cm}\bigg\{\left\Vert{\left[\begin{array}{ll}
\bi{C_{1i}}&\bi{C_{2i}}
\end{array}\right]}\right\Vert\bigg(\left\Vert\bi{\bar A_i}\right\Vert\cdot\left\Vert\bi{z}(t)\right\Vert
+r(t)\Vert\bi{\bar A_{di}}\Vert\cdot\Vert\bi{z}(t)\Vert\bigg)-\left(\epsilon_i+\Vert \bi{C_{2i}B_{2i}}\Vert f_i\right)\bigg\}\times\\
&\hspace{3cm}\mathrm{sign}\left(\bi{B_{2i}}^T\bi{C_{2i}}^T\bi{\Omega_is}(t)\right)\\
&-\frac12\left(\bi{\Omega_i}\bi{C_{2i}}\bi{B_{2i}}\right)^{-1}\sum_{j\in{\ell_\mathcal K^i}}\lambda_{ij}\left(\bi{\Omega_j}-\bi{\hat H_i}\right)\bi{s}(t)
\end{aligned}\vspace{1\fill}
\end{equation}
\centering{\rule{0.8\linewidth}{0.4pt}}
\end{figure*}
%
\begin{align}\label{rdot}
\dot r(t)&=\frac{1}{\beta}\min_{i\in N}\\
&\left\{\Vert \bi{s}(t)\Vert\Vert \bi\Omega_i\Vert\left\Vert{\left[\begin{array}{ll}
\bi{C_{1i}}&\bi{C_{2i}}
\end{array}\right]}\right\Vert\Vert\bi{\bar A_{di}}\Vert\Vert\bi{z}(t)\Vert\right\}\notag
\end{align}
where \(r(0)=0\), \(\epsilon_i>0\) is a given small constant, and \(\beta >0\) is a given scalar.
\begin{proof}\upshape
Choose the appropriate mode-dependent Lyapunov function candidate as
\begin{equation}
V(\bi z,t,i)=\frac12 \left\{\bi{s}^T(t)\bi{\Omega_is}(t)+\beta\tilde r^2(t)\right\}
\end{equation}
Applying the weak infinitesimal operator of the Lyapunov function \(\mathfrak LV(\bi z,t,i)\) and using~\eqref{s-dot}, yields
\begin{equation}\label{LV11}
\begin{aligned}
&\mathfrak LV(\bi z,t,i)=\bi s^T(t)\bi{\Omega_i}\\
&\ \quad\times\bigg\{\left[\begin{array}{ll}
\bi{C_{1i}}&\bi{C_{2i}}
\end{array}\right]\left(\bi{\bar A_iz}(t)+\bi{\bar A_{di}z}(t-\tau_i(t))\right)\bigg\}\\
&\quad +\bi s^T(t)\bi{\Omega_iC_{2i}B_{2i}}\left[\bi u(t)+\bi{F_iw}(t)\right]\\
&\quad +\frac12\sum_{j\in{\ell}}\lambda_{ij}\bi s^T(t)\bi{\Omega_js}(t)+\beta\tilde r(t)\dot{\tilde r}(t)
\end{aligned}
\end{equation}
Considering arbitrary matrices \(\bi{\hat H_i}=\bi{\hat H_i}^T\) and separating the known and unknown elements of transition probabilities matrix by (\ref{lambda}), yields:
\begin{equation}\label{LV22}
\begin{aligned}
&\mathfrak LV(\bi z,t,i)=\bi s^T(t)\bi{\Omega_i}\\
&\ \ \quad\times\bigg\{\left[\begin{array}{ll}
\bi{C_{1i}}&\bi{C_{2i}}
\end{array}\right]\left(\bi{\bar A_iz}(t)+\bi{\bar A_{di}z}(t-\tau_i(t))\right)\bigg\}\\
&\quad+\bi s^T(t)\bi{\Omega_iC_{2i}B_{2i}}[\bi u(t)+\bi{F_iw}(t)]\\
&\quad+\frac12\sum_{j\in{\ell_\mathcal K^i}}\lambda_{ij}\left(\bi s^T(t)\left(\bi{\Omega_j}-\bi{\hat H_i}\right)\bi s(t)\right)\\
&\quad +\frac12\sum_{j\in{\ell_{u\mathcal K}^i}}\lambda_{ij}\left(\bi s^T(t)\left(\bi{\Omega_j}-\bi{\hat H_i}\right)\bi s(t)\right)+\beta\tilde r(t)\dot{\tilde r}(t)
\end{aligned}
\end{equation}
Notice that, the sets of LMIs~\eqref{LMI-U11} and~\eqref{LMI-U22} are equivalent to following inequality
\begin{equation}
\frac12\sum_{j\in{\ell_{u\mathcal K}^i}}\lambda_{ij}\left(\bi s^T(t)\left(\bi{\Omega_j}-\bi{\hat H_i}\right)\bi s(t)\right)<0
\end{equation}
for  \(i\in{\ell_{\mathcal K}^i}\) and \(i\in{\ell_{u\mathcal K}^i}\), respectively. 

By considering~\eqref{raz} in Assumption~\ref{assump}, we have
\begin{equation}\label{LV333}
\begin{aligned}
\mathfrak LV(\bi z,t,i)\leq\ &\Vert\bi s(t)\Vert\cdot\Vert\bi\Omega_i\Vert\bigg\{\left\Vert\begin{bmatrix}
\bi{C_{1i}}&\bi{C_{2i}}
\end{bmatrix}\right\Vert\cdot\\
&\quad\bigg(\Vert\bi{\bar A_i}\Vert\cdot\Vert\bi z(t)\Vert +r\Vert\bi{\bar A_{di}}\Vert\cdot\Vert\bi z(t)\Vert\bigg)\bigg\}\\
&+\bi s^T(t)\bi{\Omega_iC_{2i}B_{2i}}u(t)\\
&+\left\Vert \bi{B_{2i}}^T\bi{C_{2i}}^T\bi{\Omega_is}(t)\right\Vert\cdot\left\Vert\bi{F_iw}(t)\right\Vert\\
&+\frac12\sum_{j\in{\ell_\mathcal K^i}}\lambda_{ij}\left(\bi s^T(t)\left(\bi{\Omega_j}-\bi{\hat H_i}\right)\bi s(t)\right)\\
&+\beta\tilde r(t)\dot{\tilde r}(t)
\end{aligned}
\end{equation}
By substituting SMC law~\eqref{u2} into~\eqref{LV333}, we have
\begin{multline}
\mathfrak LV(\bi z,t,i)\leq\\-\tilde{r}(t)\Vert\bi s(t)\Vert\cdot\Vert\bi{\Omega_i}\Vert\cdot\left\Vert\begin{bmatrix}
\bi{C_{1i}}&\bi{C_{2i}}
\end{bmatrix}\right\Vert\cdot\Vert\bi{\bar A_{di}}\Vert\cdot\Vert \bi z(t)\Vert \\
-\epsilon_i\Vert\bi{\Omega_i s}(t)\Vert +\beta\tilde r(t)\dot{\tilde r}(t)
\end{multline}
Now, by substituting the adaptive law~\eqref{rdot}, we have
\begin{equation}
\mathfrak LV(\bi z,t,i)\leq -\epsilon_i\Vert\bi{\Omega_i s}(t)\Vert<0
\end{equation}
where \(\epsilon_i>0\) is a given small constant. The rest of proof is similar to Theorem~\ref{thrm2} and omitted here. The proof is completed. 
\end{proof}
\end{theorem}
\section{Numerical examples}\label{sec:numerical}
In this section, we present numerical examples to illustrate the merits of the proposed approaches.
Consider the sliding mode control for system~\eqref{eq:systemxdot} with partly unknown transition probabilities~\eqref{lambda}, three operating modes, i.e. \(N=3\)  and the following system matrices and parameters:
\begin{align*}
&\bi{A_1}=\begin{bmatrix}
	-1&0\\
	2&-2
	\end{bmatrix},\bi{A_{d1}}=\begin{bmatrix}
								-2&0.1\\
								0.5&-1
								\end{bmatrix},\bi{B_1}=\begin{bmatrix}
														1\\
														0
														\end{bmatrix},\\
&\bi{A_2}=\begin{bmatrix}
	-0.15&-0.49\\
	1.5&-2.1
	\end{bmatrix},\bi{A_{d2}}=\begin{bmatrix}
								0&-3\\
								0.1&0.5
								\end{bmatrix},\\
&\bi{B_2}=\begin{bmatrix}
						2\\
						-1
			\end{bmatrix},
\bi{A_3}=\begin{bmatrix}
	-0.3&-0.15\\
	1.5&-1.8
	\end{bmatrix},\\
&\bi{A_{d3}}=\begin{bmatrix}
								-0.5&0.2\\
								0.1&-0.3
	\end{bmatrix},\bi{B_3}=\begin{bmatrix}
														1\\
														-1
														\end{bmatrix}\\
&F_1=1,\ F_2=1,\ F_3=1,\ w(t)=0.1\sin (t).
\end{align*}
The mode-dependent time-varying delay \(\tau_i(t)\) satisfies~\eqref{delay} with \(h_1=0.3,\ h_2=0.5,\ \mu_1=0.6,\ \mu_2=0.4\) and \(\mu_3=1.1\). The transition probability rate matrix is described as
\begin{align*}
	\bi\Lambda=\begin{bmatrix}
		?&?&1.1\\
		0.2&?&?\\
		0.9&0.2&-1.1
	\end{bmatrix}
\end{align*}
\paragraph{\textbf{Case I.}} By taking advantage of Matlab\textsuperscript\textcopyright \ LMI Toolbox to solve set of LMIs~\eqref{LMI1}-\eqref{LMI13} in~Theorem~\ref {thrm1}, we obtain a feasible solution as follows:
\begin{equation*}
	\begin{array}{lll}
		X_1=0.8974,\ &X_2=0.9079,\ &X_3=0.9217,\\
	Y_1=-0.1495,\ &Y_2=0.2235,\ &Y_3=1.1734,\\		
	\end{array}
\end{equation*}
and from~\eqref{Ci}, we have
\begin{equation*}
	\begin{array}{lll}
	C_1=-0.1666,&C_2=0.2462,&C_3=1.2731
	\end{array}
\end{equation*}
which give a stable sliding mode dynamics~\eqref{sl}. Now, Solving LMIs~\eqref{LMI-U1}-\eqref{LMI-U2} in Theorem~\ref {thrm2} to design a SMC law of the form~\eqref{u}, yields
\begin{equation*}
	\begin{array}{lll}
	\Omega_1=2.7392,&\Omega_2=1.4755,&\Omega_3=0.4918,\\
\hat V_1=2.1074,&\hat V_2=0.9837,&\hat V_3=1.
	\end{array}
\end{equation*}

\begin{figure}[h]%
\centering%
\includegraphics[width=\linewidth]{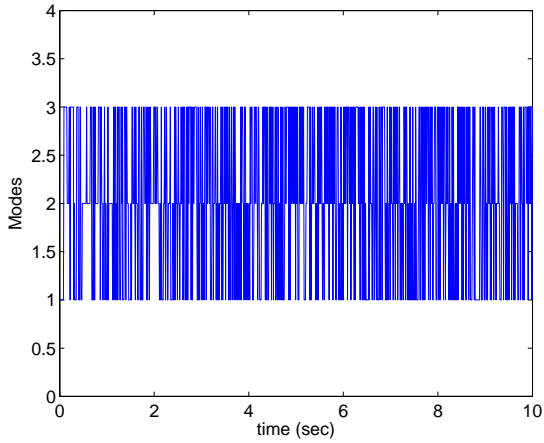}%
\caption{Markov jump state}\label{fig:mode}%
\end{figure}

\begin{figure}[ht]%
\includegraphics[width=\linewidth]{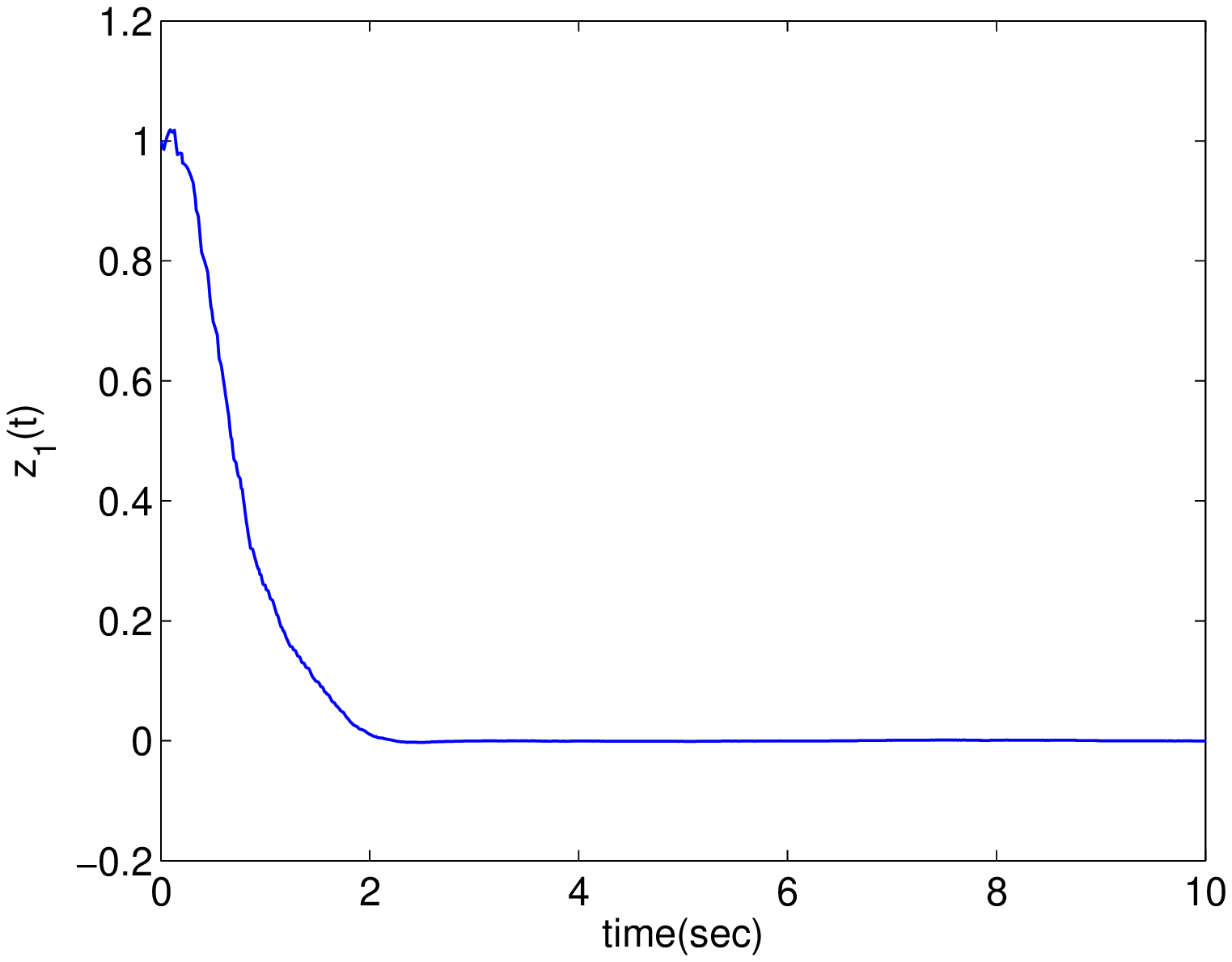}%
\caption{State vector \({z_1}(t)\) of closed loop system}\label{fig:x1}%
\end{figure}

\begin{figure}[ht]%
\centering%
\includegraphics[width=\linewidth]{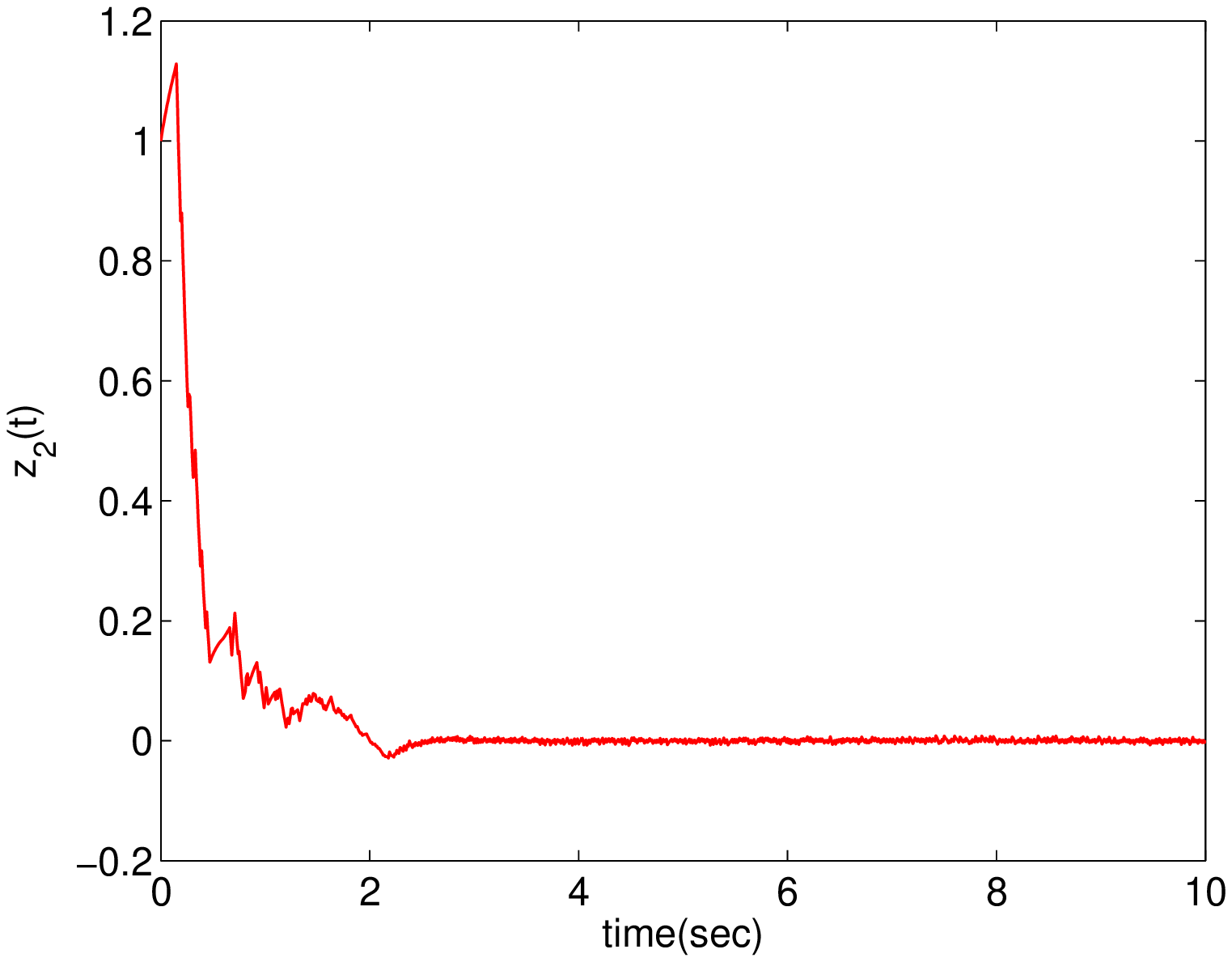}%
\caption{State vector \({z_2}(t)\) of closed loop system}\label{fig:x2}
\end{figure}

\begin{figure}[ht]%
\centering%
\includegraphics[width=\linewidth]{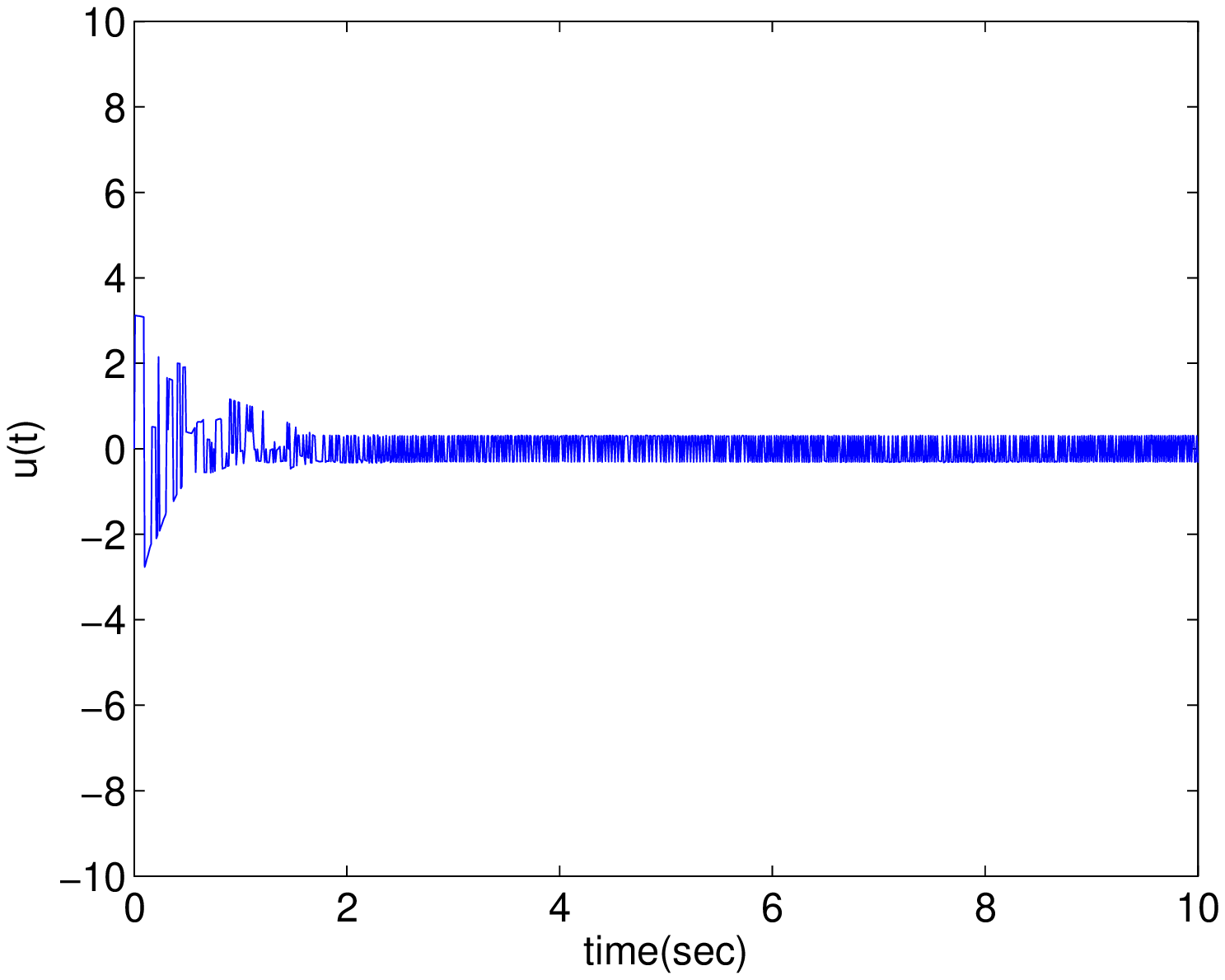}%
\caption{Control input \(u(t)\)}\label{fig:u}%
\end{figure}

\begin{figure}[ht]%
\centering%
\includegraphics[width=\linewidth]{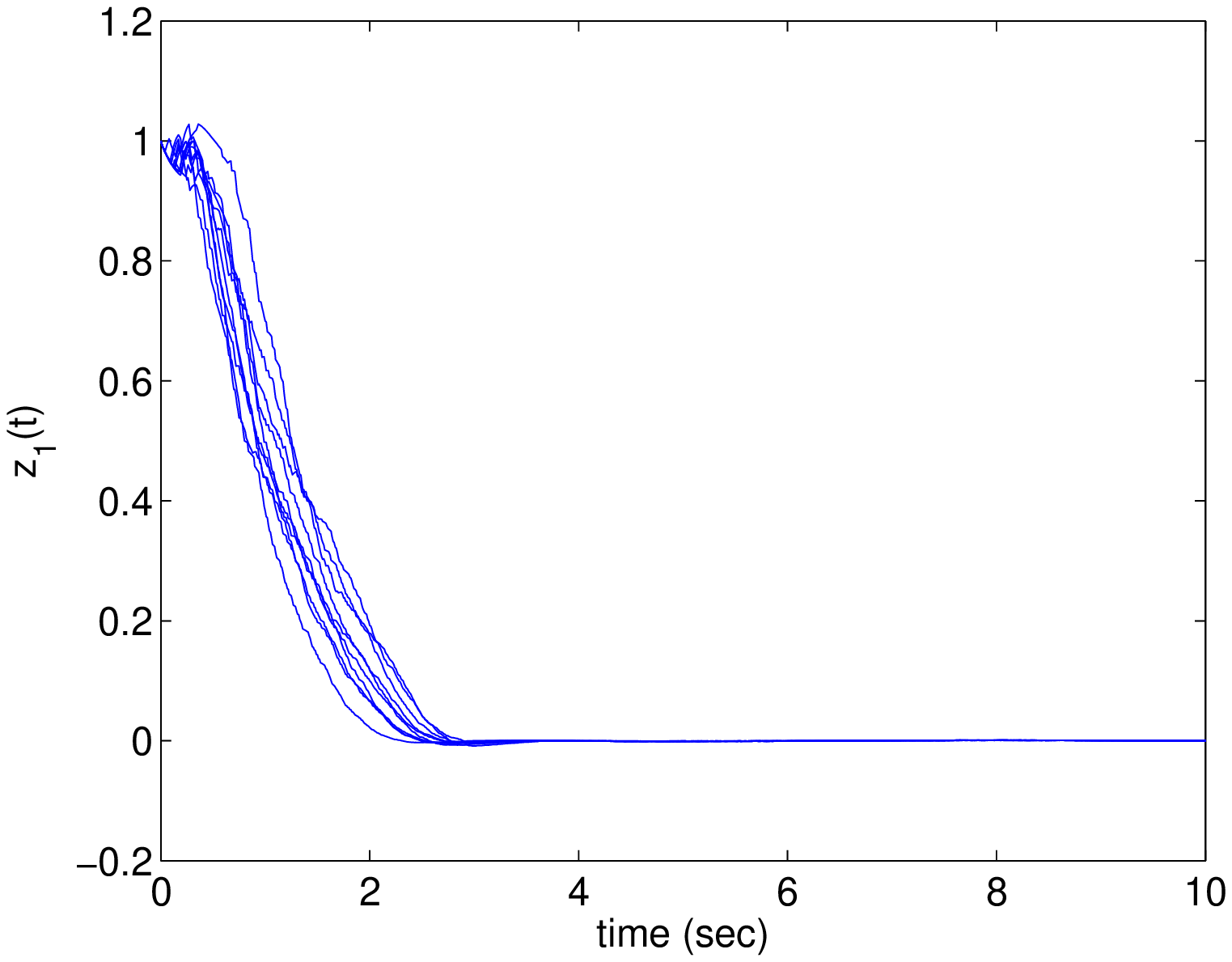}%
\caption{Simulation of 10 iterations: \(z_1(t)\)}\label{fig:10x1}
\end{figure}

\begin{figure}
\centering%
\includegraphics[width=\linewidth]{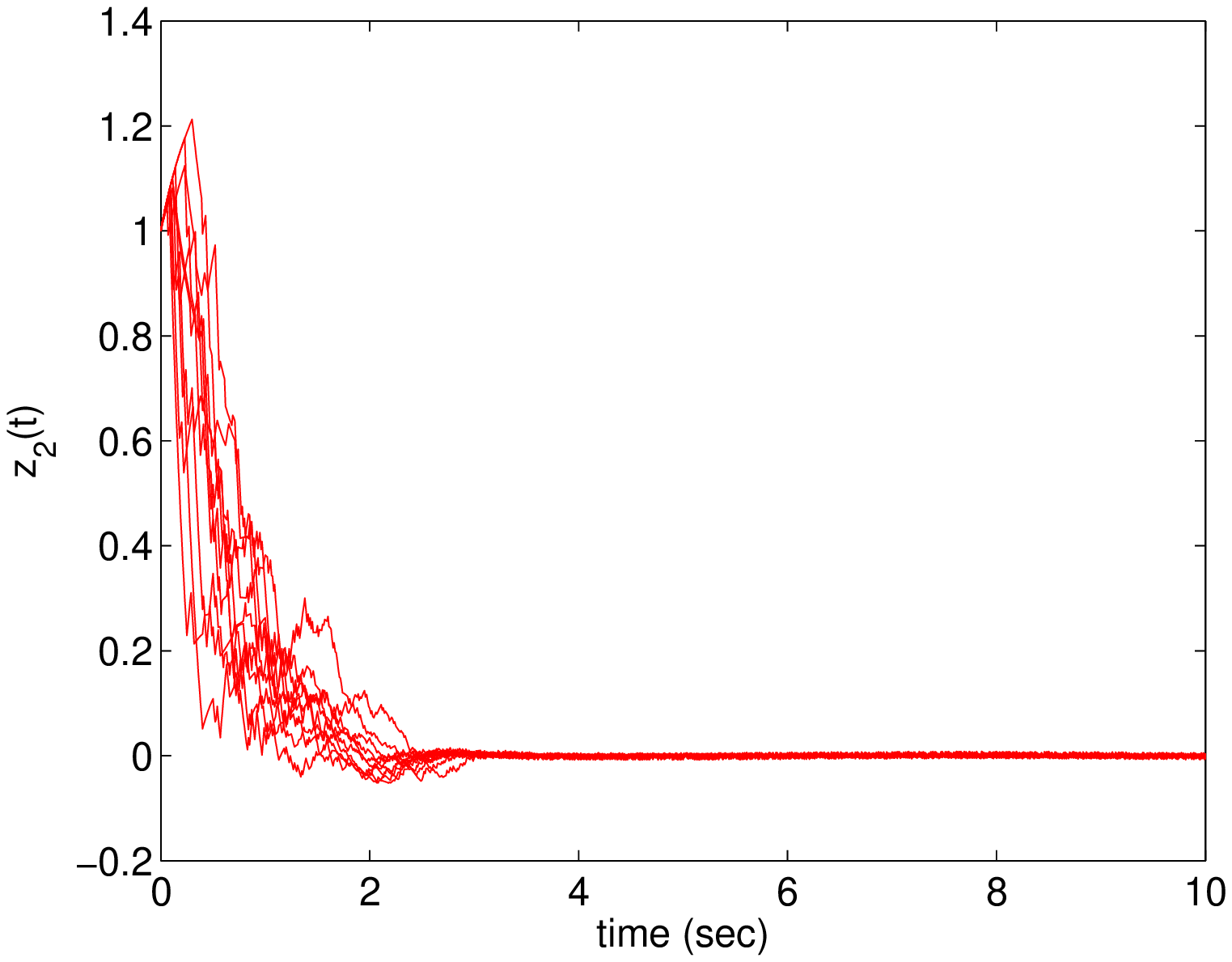}%
\caption{Simulation of 10 iterations: \(z_2(t)\)}\label{fig:10x2}
\end{figure}

By choosing \(f_1=f_2=f_3=0.1\) and \(\epsilon_1=\epsilon_2=\epsilon_3=0.2\) and considering \(\tau_1(t)=0.4+0.1 sin(5t),\tau_2(t)=0.45+0.05 sin(6t)\) and \(\tau_3(t)=0.42+0.07 cos(11t)\), we have the following simulation results: Figure~\ref{fig:mode} shows the switching of the three operating modes. Figures~\ref{fig:x1} and~\ref{fig:x2} depict the state trajectories \(\bi{z_1}(t)\)  and \(\bi{z_2}(t)\)  of the closed loop system, respectively, for the initial values \(\bi{z}(0)=\left[1\quad 1\right]^T\). Moreover, the control input \(u(t)\) is given in Figure~\ref{fig:u}. Some slight discontinuities might appear in control signal, which are effects of random jumps in Markovian jump system. To make a firm conclusion, simulation of the closed-loop system with 10 different realizations of the stochastic process \(r_t\) is done and the state trajectories \(\bi{z_1}(t)\)  and \(\bi{z_2}(t)\) are portrayed in Figures~\ref{fig:10x1} and~\ref{fig:10x2}, respectively.

\begin{figure}[ht]%
\centering%
\includegraphics[width=\linewidth]{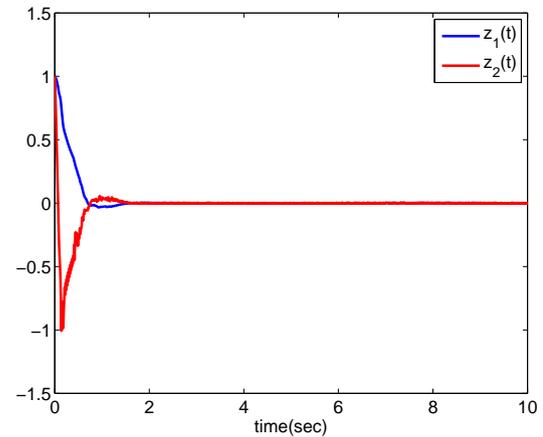}%
\caption{State vectors \(z_1(t)\) and \(z_2(t)\)}\label{fig:z1z2un}
\end{figure}

\begin{figure}[ht]
\centering%
\includegraphics[width=\linewidth]{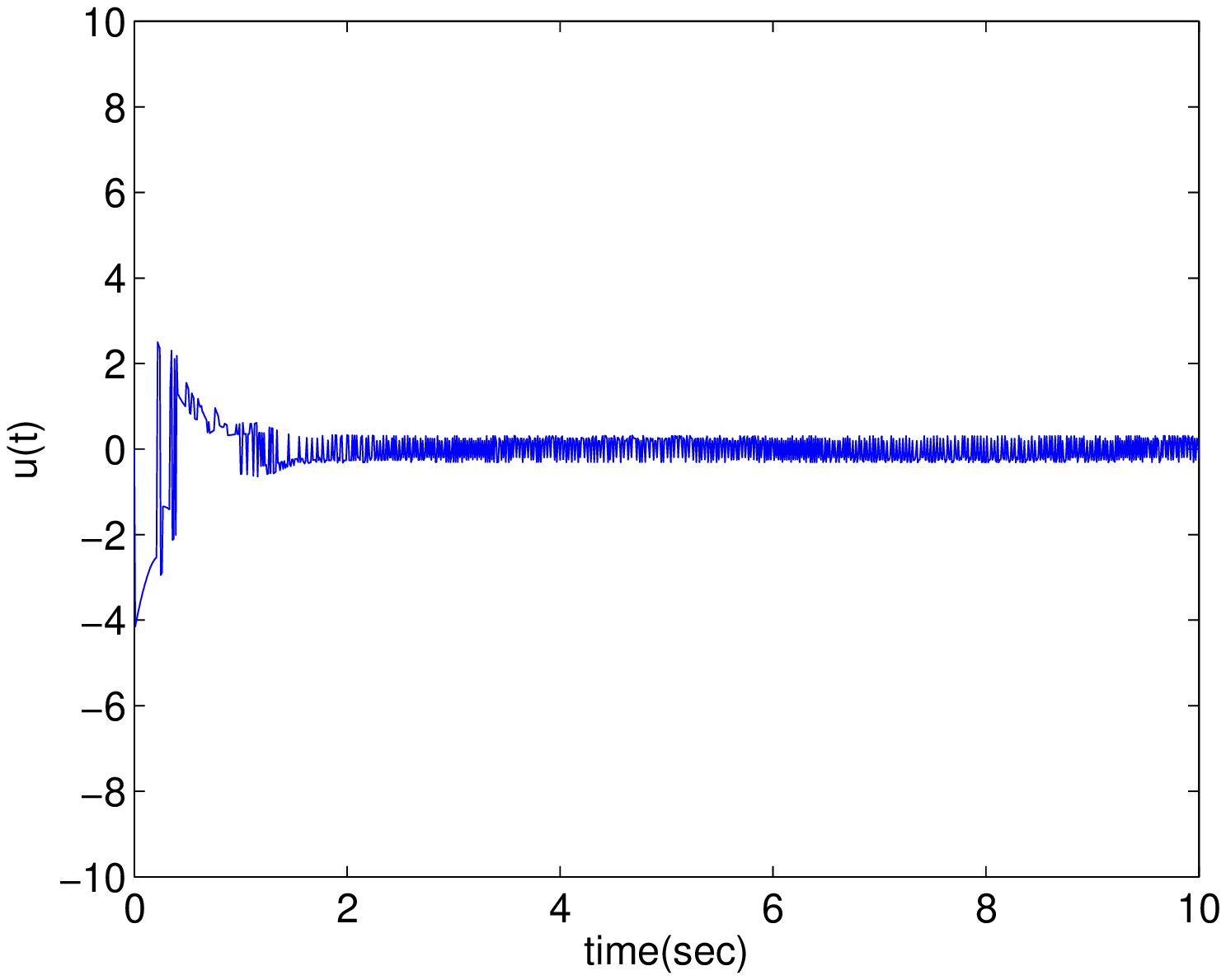}%
\caption{Control input \(u(t)\)}\label{fig:u2}
\end{figure}

\begin{figure}[ht]%
\includegraphics[width=\linewidth]{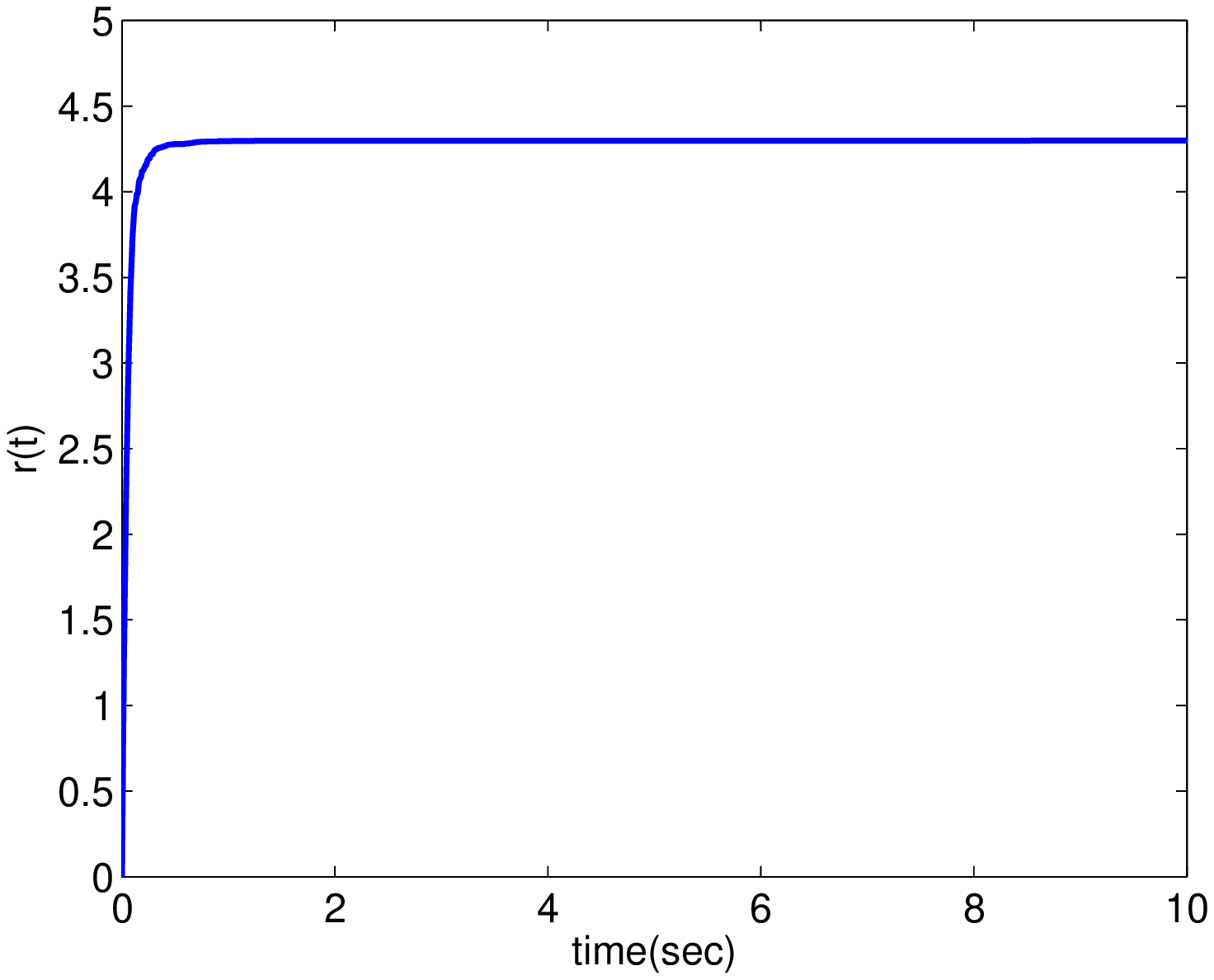}%
\caption{Adaptive Estimate \(r(t)\)}\label{fig:r(t)}
\end{figure}

\paragraph{\textbf{Case II.}}{In other situations when delay functions \(\tau_i(t)\) are unknown, by solving LMIs~\eqref{LMI-U11}-\eqref{LMI-U22} and applying the sliding mode controller~\eqref{u2}-\eqref{rdot} proposed in Theorem~\ref{thrm3}, the following simulation results are obtained: the states of the closed-loop
system \(\bi{z_1}(t)\) and \(\bi{z_2}(t)\) are shown in Figure~\ref{fig:z1z2un} with the initial values given by \(\bi{z}(0)=\left[1\quad 1\right]^T\). Moreover, the control input \(u(t)\), and  \(r(t)\) with initial condition \(r(0)=0\) are portrayed in Figures~\ref{fig:u2} and~\ref{fig:r(t)}, respectively.
The adaptive law is given as:
\begin{equation}
\dot r(t)=0.1454\ \Vert\bi s(t)\Vert\cdot\Vert\bi z(t)\Vert
\end{equation}
with \(\beta=2\).}

The simulation results demonstrate that by applying the proposed SMC law, the state trajectories of the closed-loop system are driven onto the predefined sliding surface in finite time which verifies our main results.
\section{Conclusion}\label{sec:conc}
A sliding mode control design for mode-dependent time varying delayed Markovian jump systems with partly unknown transition probabilities has been investigated. By using a new stochastic Lyapunov Krasovskii functional candidate combining with Jensen's inequality and free-connection weighting matrix method, the sufficient delay-dependent conditions for stochastic stability of sliding mode dynamics has been presented in terms of LMIs. A SMC law has been synthesized to ensure the reachability of the closed-loop system's state trajectories to the specified sliding surface in finite time. In our design approach, the information of delay size has been considered and derivative of mode-dependent time-varying delay may be larger than one. Therefore, less-conservative criteria have been derived. {In addition, an adaptive sliding mode controller has been designed to apply to cases, where mode-dependent time-varying delays are unknown.} All of the conditions for the stability of sliding mode dynamics and SMC law design are expressed in terms of LMIs. Finally, numerical examples have been provided to demonstrate the validity of the main results.



\begin{thebibliography}{10}

\bibitem{bib33}
Krasovskii NN, Lidskii EA.
\newblock Analytical design of controllers in systems with random attributes.
  I. Statement of the problem, method of solving.
\newblock Automatic Remote Control. 1961;22:1021 -- 1294.

\bibitem{bib2}
Boukas EK, Zhang Q, Yin G.
\newblock Robust production and maintenance planning in stochastic
  manufacturing systems.
\newblock IEEE Transactions on Automatic Control. 1995;40(6):1098--1102.

\bibitem{bib3}
Narendra KS, Tripathi SS.
\newblock Identification and Optimization of Aircraft Dynamics.
\newblock J Aircraft. 1973;10(4):193--199.

\bibitem{bib4}
Mao Z, Jiang B, Shi P.
\newblock \hinf fault detection filter design for networked control systems
  modelled by discrete {Markovian} jump systems.
\newblock IET Control Theory and Applications. 2007;1(5):1336--1343.

\bibitem{bib5}
Zohrabi N, Momeni HR, Zakeri H.
\newblock {Markov} Jump Modelling and Control of Stress Effects on the Blood
  Glucose Regulation System.
  \newblock In the 20th Iranian Conference on Electrical Engineering (ICEE2012), pp. 985--989, 2012. 
  
  \bibitem{jadid1}
  Shi P, and Li F. A survey on Markovian jump systems: modeling and design. International Journal of Control, Automation and Systems, vol. 13, no. 1 (2015): 1--16.
  
\bibitem{Z3}
Zohrabi N, Shi J, Abdelwahed S. Ship-wide transient specifications and criteria for Medium-Voltage DC Shipboard Power System. 2016 IEEE Transportation Electrification Conference and Expo (ITEC), 2016 IEEE, pp. 1--6, 2016.

\bibitem{Z4}
Zohrabi N, Shi J,  Abdelwahed S. Steady State Specifications and Requirements for Medium-Voltage DC Shipboard Power System. ASNE Advanced Machinery Technology Symposium, Philadelphia, May 2016.

\bibitem{bib7}
Boukas EK.
\newblock Stochastic Switching Systems: Analysis and Design.
\newblock Control Engineering. Birkh{\"a}user; 2006.




\bibitem{bib100}
Boukas EK.
\newblock On stability and stabilisation of continuous time singular
  {Markovian} switching systems.
\newblock Control Theory Applications, IET. 2008; 2(10):884--894.

\bibitem{bib101}
Lin Z, Lin Y, Zhang W.
\newblock \hinf filtering for non-linear stochastic {Markovian} jump systems.
\newblock Control Theory Applications, IET. 2010; 4(12):2743--2756.

\bibitem{bib11}
Shi P, Boukas EK, Agarwal K.
\newblock Kalman filtering for continuous-time uncertain systems with
  {Markovian} jumping parameters.
\newblock IEEE Transactions on Automatic Control. 1999; 44(8):1592--1597.

\bibitem{jadid2}
Park BY, Kwon NK, Park P. Stabilization of Markovian jump systems with incomplete knowledge of transition probabilities and input quantization. Journal of the Franklin Institute. 2015 Oct 31;352(10):4354--65.
\bibitem{bib13}
Zhang LX, Boukas EK.
\newblock Stability and stabilization of {Markovian} jump linear systems with
  partly unknown transition probabilities.
\newblock Automatica. 2009;45(2):463--468.

\bibitem{bib103}
Shen M, Yang GH.
\newblock New analysis and synthesis conditions for continuous Markov jump
  linear systems with partly known transition probabilities.
\newblock Control Theory Applications, IET. 2012;6(14):2318--2325.

\bibitem{bib19}
Luan X, Liu F, Shi P.
\newblock Finite-time filtering for non-linear stochastic systems with
  partially known transition jump rates.
\newblock Control Theory and Applications, IET. 2010;4(5):735--745.

\bibitem{bib21}
Zhang L, Lam J.
\newblock Necessary and sufficient conditions for analysis and synthesis of
  {Markov} jump linear systems with incomplete transition descriptions.
\newblock IEEE Transactions on Automatic Control. 2010;55(7):1695--1701.

\bibitem{bib49}
Boukas EK.
\newblock Control of stochastic systems with time-varying multiple time delays:
  {LMI} approach.
\newblock Journal of optimization theory and applications. 2003;119(1):19--36.

\bibitem{bib50}
Boukas EK, Liu ZK.
\newblock Deterministic and stochastic time delay systems.
\newblock Boston: Birkh{\"a}user; 2002.

\bibitem{bib51}
Mohammadian M, Hossein~Abolmasoumi A, Reza~Momeni H.
\newblock \hinf mode-independent filter design for Markovian jump genetic
  regulatory networks with time-varying delays.
\newblock Neurocomputing. 2012;p. 10--18.

\bibitem{bib52}
Abolmasoumi AH, Momeni HR.
\newblock Robust observer-based \hinf control of a Markovian jump system with
  different delay and system modes.
\newblock International Journal of Control, Automation and Systems.
  2011;9(4):768--776.
  
\bibitem{newbib1}
Liu X, Xi H. Stochastic stability for uncertain neutral Markovian jump systems with nonlinear perturbations. Journal of Dynamical and Control Systems. 21, no. 2 (2015): 285--305.

\bibitem{bib56}
Boukas EK, Liu Z.
\newblock Robust stability and stabilizability of Markov jump linear uncertain
  systems with mode-dependent time delays.
\newblock Journal of Optimization Theory and Applications.
  2001;109(3):587--600.

\bibitem{bib55}
Xu S, Chen T, Lam J.
\newblock Robust \hinf filtering for uncertain Markovian jump systems with
  mode-dependent time delays.
\newblock IEEE Transactions on Automatic Control. 2003;48(5):900--907.

\bibitem{bib102}
Ma S, Boukas EK.
\newblock Robust \hinf filtering for uncertain discrete Markov jump singular
  systems with mode-dependent time delay.
\newblock Control Theory Applications, IET. 2009;3(3):351--361.

\bibitem{jadid3}
Ren J, Zhu H, Zhong S, Zhou X. Robust stability of uncertain Markovian jump neural networks with mode-dependent time-varying delays and nonlinear perturbations. Advances in Difference Equations. 2016 Dec 15;2016(1):327.
\bibitem{bib23}
Shi P, Xia Y, Liu GP, Rees D.
\newblock On designing of sliding-mode control for stochastic jump systems.
\newblock IEEE Transactions on Automatic Control. 2006;51(1):97--103.

\bibitem{bib24}
Niu Y, Ho DWC, Wang X.
\newblock Sliding mode control for {It\^o} stochastic systems with {Markovian}
  switching.
\newblock Automatica. 2007;43(10):1784--1790.

\bibitem{jadid4}
Li H, Shi P, Yao D, Wu L. Observer-based adaptive sliding mode control for nonlinear Markovian jump systems. Automatica. 2016 Feb 29;64:133--42.

\bibitem{bib25}
Wu L, Shi P, Gao H.
\newblock State estimation and sliding-mode control of {Markovian} jump
  singular systems.
\newblock IEEE Transactions on Automatic Control. 2010;55(5):1213--1219.

\bibitem{bib26}
Zhang Y, Xu S, Chu Y.
\newblock Sliding mode observer-controller design for uncertain {Markovian}
  jump systems with time delays.
\newblock International Journal of Robust and Nonlinear Control.
  2012;22(4):355--368.

\bibitem{Z1}
Zohrabi N, Momeni HR, Abolmasoumi AH.
\newblock Sliding Mode Control of Time-Delay Markovian Jump Systems with Partly
Known Transition Probabilities. Journal of Control, Vol. 6, No. 3, pp. 61--70 ,2012.

\bibitem{Z2}
Zohrabi N, Abolmasoumi AH, Momeni HR, Abdelwahed S.
\newblock Robust Integral Sliding Mode Control for Markovian Jump Systems: A Singular System Approach. In 2016 American Control Conference (ACC), pp. 7213-7218, 2016.


\bibitem{IFAC}
Zohrabi N, Momeni HR, Abolmasoumi AH.
\newblock Sliding Mode Control of Markovian Jump Systems with Partly Unknown
  Transition Probabilities.
\newblock IFAC Proceedings, Vol. 46, no. 2, pp. 947--952, 2013.

\bibitem{bib555}
Niu Y, Ho DWC, Lam J.
\newblock Robust integral sliding mode control for uncertain stochastic systems
  with time-varying delay.
\newblock Automatica. 2005;41(5):873--880.

\bibitem{bib666}
Wu L, Su X, Shi P.
\newblock Sliding mode control with bounded \ensuremath{\mathscr{L}_2} gain
  performance of Markovian jump singular time-delay systems.
\newblock Automatica. 2012;48(8):1929--1933.

\bibitem{bib34}
Xu S, Chen T.
\newblock An {LMI} approach to the \hinf filter design for uncertain systems
  with distributed delays.
\newblock IEEE Transactionson Circuits and Systems. 2004;51(4):195--201.

\bibitem{bib37}
Yue D, Tian E, Zhang Y.
\newblock A piecewise analysis method to stability analysis of linear
  continuous/discrete systems with time-varying delay.
\newblock International Journal of Robust Nonlinear Control.
  2009;19(13):1493--1518.

\bibitem{bib28}
Hung J, Gao W, Hung JC.
\newblock Variable structure control: A survey.
\newblock IEEE Transactions on Industrial Electronics. 1993;40(1):2--22.

\bibitem{bib29}
Xia Y, Jia Y.
\newblock Robust sliding-mode control for uncertain time-delay systems: An
  {LMI} approach.
\newblock IEEE Transactions on Automatic Control. 2003;48(6):1086--1092.

\bibitem{bib30}
Kushner HJ.
\newblock Stochastic Stability and Control.
\newblock Mathematics in Science and Engineering. Academic Press; 1967.

\bibitem{bib38}
Gu K.
\newblock An intergral inequality in stability problem of time-delay systems.
\newblock Proceeding of 39th IEEE Conference on Decision and Control.
  2000;3:2805--2810.

\bibitem{bib41}
He S, Liu F.
\newblock Observer-based finite-time control of time-delayed jump systems.
\newblock Applied Mathematics and Computation. 2010;217(6):2327--2338.

\bibitem{bib42}
Wen Jw, Liu F, Kiong~Nguang S.
\newblock Sampled-data predictive control for uncertain jump systems with
  partly unknown jump rates and time-varying delay.
\newblock Journal of the Franklin Institute. 2012;349(1):305--322.

\bibitem{bib43}
Shao H.
\newblock Delay-range-dependent robust \hinf filtering for uncertain stochastic
  systems with mode-dependent time delays and Markovian jump parameters.
\newblock Journal of Mathematical Analysis and Applications.
  2008;342(2):1084--1095.

\bibitem{bib44}
Wang Z, Huang L, Yang X.
\newblock \hinf performance for a class of uncertain stochastic nonlinear
  Markovian jump systems with time-varying delay via adaptive control method.
\newblock Applied Mathematical Modelling. 2011;35(4):1983--1993.

\bibitem{bib47}
Boukas EK.
\newblock Control of stochastic systems with time-varying multiple time delays:
  {LMI} approach.
\newblock Journal of optimization theory and applications. 2003;119(1):19--36.

\bibitem{bib46}
Fei Z, Gao H, Shi P.
\newblock New results on stabilization of Markovian jump systems with time
  delay.
\newblock Automatica. 2009;45(10):2300--2306.

\bibitem{bib45}
Li Z, Wang S, Fei Z, Gao H.
\newblock Further results on \hinf control for Markovian jump systems with
  mode-dependent time-varying delays.
\newblock In 3rd International Symposium on Systems and Control in Aeronautics
  and Astronautics; 2010. p. 1521--1526.
  
\bibitem{comment}
Zohrabi N, Momeni HR, Abolmasoumi AH. Comments on “Delayed-state-feedback exponential stabilization for uncertain Markovian jump systems with mode-dependent time-varying state delays”. Nonlinear Dynamics. vol. 73, no. 3, pp. 1493--4, 2013.

\bibitem{bib48}
Kushner HJ.
\newblock Stochastic stability and control.
\newblock Mathematics in Science and Engineering. Elsevier Science; 1967.

\bibitem{bib777}
Hale JK, Lunel SMV.
\newblock Introduction to Functional Differential Equations.
\newblock Springer: New York; 1993.

\end{thebibliography}
\end{document}